\documentclass[twocolumn]{article}
\usepackage{cite}
\usepackage[utf8]{inputenc}
\usepackage[english]{babel}
\usepackage[T1]{fontenc}
\usepackage{amsmath}
\usepackage{amssymb}
\usepackage[standard]{ntheorem}
\usepackage{stmaryrd}
\usepackage{xcolor}
\usepackage{enumitem}
\usepackage{graphicx}
\usepackage{geometry}
\usepackage{csquotes}
\usepackage{enumitem}
\usepackage{algorithm} 
\usepackage{algorithmic} 
\usepackage{microtype}
\usepackage[utf8]{inputenc} 
\usepackage{hyperref}       
\usepackage{booktabs}       
\usepackage{multirow} 
\usepackage{subfigure}
\usepackage{amsfonts,amsmath,amssymb,color}

\usepackage{helvet} 
\usepackage{courier} 
\usepackage{color}

\newcommand{\Cmat}[0]{\ensuremath{{\bf C}} }
\newcommand{\Dmat}[0]{\ensuremath{{\bf D}} }

\newcommand{\Fmat}[0]{\ensuremath{{\bf F}} }
\newcommand{\Gmat}[0]{\ensuremath{{\bf G}} }
\newcommand{\Hmat}[0]{\ensuremath{{\bf H}} }

\newcommand{\Xmat}[0]{\ensuremath{{\bf X}} }
\newcommand{\Ymat}[0]{\ensuremath{{\bf Y}} }

\newcommand{\gv}[0]{\ensuremath{\boldsymbol{g}} }

\newcommand{\xv}[0]{\ensuremath{\boldsymbol{x}} }
\newcommand{\yv}[0]{\ensuremath{\boldsymbol{y}} }

\newcommand{\Thetamat}[0]{\ensuremath{\boldsymbol{\Theta}} }

\newcommand{\Phimat}[0]{\ensuremath{\boldsymbol{\Phi}}}

\newcommand{\sigmav}[0]{\ensuremath{\boldsymbol{\sigma}} }

\newcommand{\ie}{{\em i.e.}}
\newcommand{\eg}{{\em e.g.}}
\newcommand{\etc}{{\em etc.}}

\newcommand{\argmin}{\arg\min}

\title{Dual-view Snapshot Compressive Imaging via Optical Flow Aided Recurrent Neural Network}

\author{\protect\textsc{Ruiying Lu, Bo Chen \thanks{Correspoding author: bchen@mail.xidian.edu.cn.}, Guanliang Liu, Ziheng Cheng},
\\ \normalsize National Lab of Radar Signal Processing, Xidian University \\
\protect\textsc{Mu Qiao}, \\ \normalsize School of Physical Science and Technology, Ningbo University \\
\protect\textsc{Xin Yuan \thanks{Correspoding author: xyuan@westlake.edu.cn.}} \\ \normalsize Westlake University
}
\date{\today}
\begin{document}

\maketitle

\maketitle
\begin{abstract}
Dual-view snapshot compressive imaging (SCI) aims to capture videos from two field-of-views (FoVs) using a 2D sensor (detector) in a single snapshot, achieving joint FoV and temporal compressive sensing, and thus enjoying the advantages of low-bandwidth, low-power and low-cost. However, it is challenging for existing model-based decoding algorithms to reconstruct each individual scene, which usually require exhaustive parameter tuning with extremely long running time for large scale data. In this paper, we propose an optical flow-aided recurrent neural network for dual video SCI systems, which provides high-quality decoding in seconds. Firstly, we develop a diversity amplification method to enlarge the differences between scenes of two FoVs, and design a deep convolutional neural network with dual branches to separate different scenes from the single measurement. {Secondly, we integrate the bidirectional optical flow extracted from adjacent frames with the recurrent neural network to jointly reconstruct each video in a sequential manner.} Extensive results on both simulation and real data demonstrate the superior performance of our proposed model in short inference time. {The code and data are available at \url{https://github.com/RuiyingLu/OFaNet-for-Dual-view-SCI}}.
\end{abstract}
{}
\section{Introduction}
As a well-developed technique, compressive sensing (CS) \cite{Donoho06ITT,Candes06ITT} based computational imaging systems \cite{Mait18CI} have attracted much attention in recent years. These systems usually employ low-dimensional sensors to capture high-dimensional signals. Snapshot compressive imaging (SCI)~\cite{Yuan2021_SPM} is one of the promising applications of computational imaging \cite{Patrick13OE,Wagadarikar08CASSI}, which utilizes a two dimensional (2D) camera to capture the 3D video or spectral data, aiming to provide an effective optical encoder for compressing the video or spectral data-cube during capture. Different from conventional cameras, such imaging systems encode the consecutive video frames \cite{reddy2011p2c2,Yuan14CVPR} or spectral channels \cite{Wagadarikar09CASSI,Miao19ICCV} by applying different coding patterns to each data volume along time or spectrum to obtain the final compressed measurements. With such cameras, SCI systems \cite{Hitomi11ICCV,Patrick13OE,reddy2011p2c2,Wagadarikar08CASSI,Wagadarikar09CASSI} can encode multiple frames into a single captured image (called measurement in this work) with low-memory, low-bandwidth, low-power and potentially low-cost, which is decoded later using different reconstruction algorithms.

While previous video SCI systems only capture one scene within the field-of-view (FoV) through spatial multiplexing, extending the spatial multiplexing capability to more than one additional dimensions, \ie, multi-view imaging, is able to decrease the memory, bandwidth as well as the latency \cite{qiao2020snapshot},
{which is necessarily required for the emerging applications such as robotics, traffic surveillance, sports photography and self-driving cars \cite{Lu20SEC,qiao2020snapshot}}.
Recently, the snapshot spatial-temporal compressive imaging (SSTCI) system implemented in \cite{qiao2020snapshot} has achieved spatial (joint FoV) and temporal CS in a snapshot by using a single coding device, \eg, the digital micro-mirror device (DMD), plus elegant simple optical designs without compromising spatial resolution, \ie, each FoV is of full resolution. The underlying principle of SSTCI is masking various scenes with different coding patterns along time to obtain the final compressed measurements. This joint CS sheds lights on the real applications of SCI system on robotics since {\em multi-view} is an inherent requirement in machine vision.
For efficient dual-view video reconstruction, a plug-and-play framework with deep denoising priors (PnP-TV-FFD) was proposed in \cite{qiao2020snapshot}. {However, one bottleneck of this method is the slow reconstruction speed and poor reconstruction quality that preclude the wide applications of SSTCI.
Inspired by this novel hardware design and aiming to address the challenges of reconstruction speed and quality, we intend to develop an effective and efficient reconstruction algorithm for SSTCI. Though our long-term goal is {\em multi-view}, in this paper, we focus on the {\em dual-view} case based on the SSTCI prototype without the loss of generality.}

For fast inference, one straightforward way is to train an end-to-end network based on deep learning for the inversion problem \cite{Cheng20ECCV_BIRNAT,Ma19ICCV,Qiao2020_APLP}. Yet, these networks were developed for the single-view SCI, and extending the single-view SCI inversion techniques to the muti-view reconstruction directly is non-trivial because they usually fail to clearly reconstruct each individual scene from the single compressed measurement. 
{It is worth noting that in the SSTCI hardware design, though }the coding pattern sets are different for the two FoVs, they are correlated as one set is a shifted version of the other one~\cite{qiao2020snapshot}. Inspired by this, we propose a {\em diversity amplifier} in this paper to facilitate the distinction of two scenes by taking the different (shifted) coding patterns into consideration, which is implemented in conjunction with a {\em dual-net separator} with two branches for reconstructing each video frames respectively. By this novel design, the two FoV scenes can be well distinguished. 
{Nevertheless, it neglects the inherent temporal correlation within each video, leading to limited reconstruction quality.} To address this, we further introduce {\em optical flow to understand the video contents for vision tasks} in our network design, which is able to investigate the motion information by encoding the velocity and direction of each pixel's movement between neighboring frames.
Furthermore, the extracted optical flow are explored bidirectionally and then fed into a recurrent neural network (RNN) to exploit temporal correlations between adjacent frames, resulting in reconstructed videos with more smoothness across time and less artifacts. It is essential to know both the object status (e.g., captured by the RNN cell) and motion information (e.g., optical flow), for better understanding the video contents of vision tasks.

\subsection{Contributions of This Paper}
In a nutshell, we develop an {\em Optical Flow-aided Recurrent Neural Network} (OFaNet) for dual-view video SCI. Specific contributions are summarized as follows:
\begin{itemize}
    \item {An end-to-end deep learning based reconstruction regime is proposed for dual-view video SCI reconstruction, where a {\em diversity amplifier} and {\em dual-net separator} are constructed to separate the two FoVs from a single measurement, and then a {\em refine net} equipped with the bidirectional optical flow is developed to improve the reconstruction quality {in a sequential manner}.}
    \item {The {\em bidirectional optical flow} is elaborately integrated into the video SCI by exploiting explicit motion information between adjacent frames, including both the global and locally varying motions, acting as the guidance for improving video reconstruction. To the best of our knowledge, this is the first time that optical flow is introduced into video SCI problems, which is jointly optimized under the supervision of mean square error (MSE) loss of reconstruction results to better match the video SCI problem.}
    \item We apply our developed algorithms to extensive simulation and real datasets (captured by the SSTCI camera) to verify the effectiveness, efficiency and robustness of the proposed algorithm. {{Experimental} results show superior performance and robustness with much shorter inference time (about 40,000 times speedups over DeSCI) compared with previous state-of-the-art methods.}
    \item {We also extend the proposed OFaNet to the single-view SCI system on both simulation and real datasets with small modifications to enlarge the application scope. The reconstruction performance is comparable with the state-of-the-art methods for single-view video SCI.}
\end{itemize}

\subsection{Related Work}

\subsubsection{Single-view SCI}
For temporal CS, SCI system is built as an {\em optical encoder}, by compressing the videos during capture. The modulation methods can be categorized into physical mask~\cite{Patrick13OE,Yuan14CVPR} and spatial light modulator (SLM) including DMD~\cite{Qiao2020_APLP,reddy2011p2c2,Sun17OE}. Given the measurement and coding patterns, an efficient decoder is required for reconstructing the time series of video frames, which has been extensively developed before \cite{Yuan16ICIP_GAP,Liu18TPAMI,YangYLLBSC14,2016arXivVideoCS,Yuan20PnPSCI,PnP_SCI_arxiv2021,Meng_GAPnet_arxiv2020}.

For the {\em software decoder}, iterative optimization based methods were proposed including GAP-TV~\cite{Yuan16ICIP_GAP,Yuan_TVSCI_arxiv2020}, GMM \cite{YangYLLBSC14}, DeSCI~\cite{Liu18TPAMI}, utilizing different priors such as total variation, low rank, sparsity prior for video SCI but suffering from either the slow reconstruction speed or poor reconstruction quality. Recently, several deep learning based methods have achieved performances comparable with traditional methods while being significantly faster at the inference stage. {Specifically}, the deep learning based approaches~\cite{Qiao2020_APLP} have been explored such as fully connected network based method~\cite{Iliadis18DSPvideoCS}, end-to-end convolutional neural network (CNN) U-net~\cite{Qiao2020_APLP}, deep tensor based ADMM-net~\cite{Ma19ICCV}, hardware constraints based deep neural network (DNN)~\cite{Yoshida18ECCV}, RNN-based BIRNAT~\cite{Cheng20ECCV_BIRNAT}, to reconstruct video images with higher quality and higher speed. In this paper, encouraged by those works, we develop an end-to-end framework for joint spatial and temporal compressive imaging reconstruction with an exemplar system developed in~\cite{qiao2020snapshot}.

\subsubsection{Spatial and temporal SCI}
{Instead of deploying multiple video SCI cameras by increasing the memory, bandwidth as well as the latency, the {\em joint spatial (multi-view) and temporal compressive imaging} provides a low-cost, low-bandwidth and low-memory requirement imaging system by capturing high-speed motions of multi views within a snapshot,} which is in demand for the emerging applications such as traffic surveillance, sports photography and self-driving cars \cite{Lu20SEC,qiao2020snapshot}. For example, for some important vision tasks such as autonomous navigation, site modelling and surveillance for robotics, conventional cameras with a limited FoV often lose sight of objects when their bearings change suddenly due to a significant turn of the observer (\eg, a robot), the object, or both \cite{SPACEK20053}. In contrast, multi-view sensors can track objects more robustly meanwhile enjoying the advantages of low cost and feasibility in hardware.
For spatial (not limited to FoV) compressive imaging, besides modulating each scene with a shifted version of the whole aperture as used in SSTCI, another intuitive method is partitioning a coded aperture to different blocks and then merging them together, but this method is usually too expensive or even infeasible to realize in hardware.
{Moreover}, a quantization with entropy coding approach based on frame skipping was adopted in~\cite{AngayarkanniRA19} to achieve efficient multi-view wireless video compression based on CS.
In addition, a lensless camera was proposed in~\cite{NakamuraKTY19} with a novel design to capture the front and back sides simultaneously, where the object-side sensor works as a coding mask and the other one as a sparsified image sensor.
In summary, storage and transmission of multi-view video sequences involve large volumes of redundant data, which can be efficiently compressed by the SSTCI system (hardware) and then resolved using a decoder (software).
By performing the compressive sensing-based image reconstruction, the captured coded image can be decoded computationally. This paper aims to develop an efficient and effective end-to-end deep learning based decoder for multi-view video SCI reconstruction. 

\subsubsection{Optical Flow}
Optical flow is a long-standing vision task aiming to estimate the per-pixel motion between video frames, which provides a plausible motion information facilitating lots of downstream
tasks such as video alignment \cite{Compensation}, video editing \cite{Inpainting}, and video
analysis \cite{ChengTW017,Beyond}.
Optical flow estimation has traditionally been treated as an energy minimization problem and recently shows a promising alternative trend by deep learning methods being significantly faster at inference stage.
One of the milestone works of deep learning for optical flow estimation is FlowNet~\cite{FlowNet}, which is based on an U-net auto-encoder {and achieves promising performance}.
Following this work, more architectures for optical flow estimation have been evolved in recent years, yielding better results with faster inference time, such as FlowNet2~\cite{IlgMSKDB17}, PWCNet~\cite{PWCNet} and LiteFlowNet~\cite{LiteFlowNet}, RAFT~\cite{RAFT}. These methods typically adopt an iterative updating approach training with the synthetic datasets and involve operators like cost volume, pyramidal features, and backward feature warping.
Taking both the efficiency and accuracy into consideration, we employ the FlowNet2~\cite{IlgMSKDB17} as our optical flow estimator, while the optical flow is fed into the recurrent neural network for SCI reconstruction as the explicit motion guidance, which has not been studied by any {previous} video SCI approaches.

\section{Review of the  Snapshot Spatial-Temporal Compressive Imaging System in~\cite{qiao2020snapshot}}
\subsection{Hardware Principle}
\begin{figure}[!htbp]
	\centering
    \includegraphics[width=1.0\columnwidth]{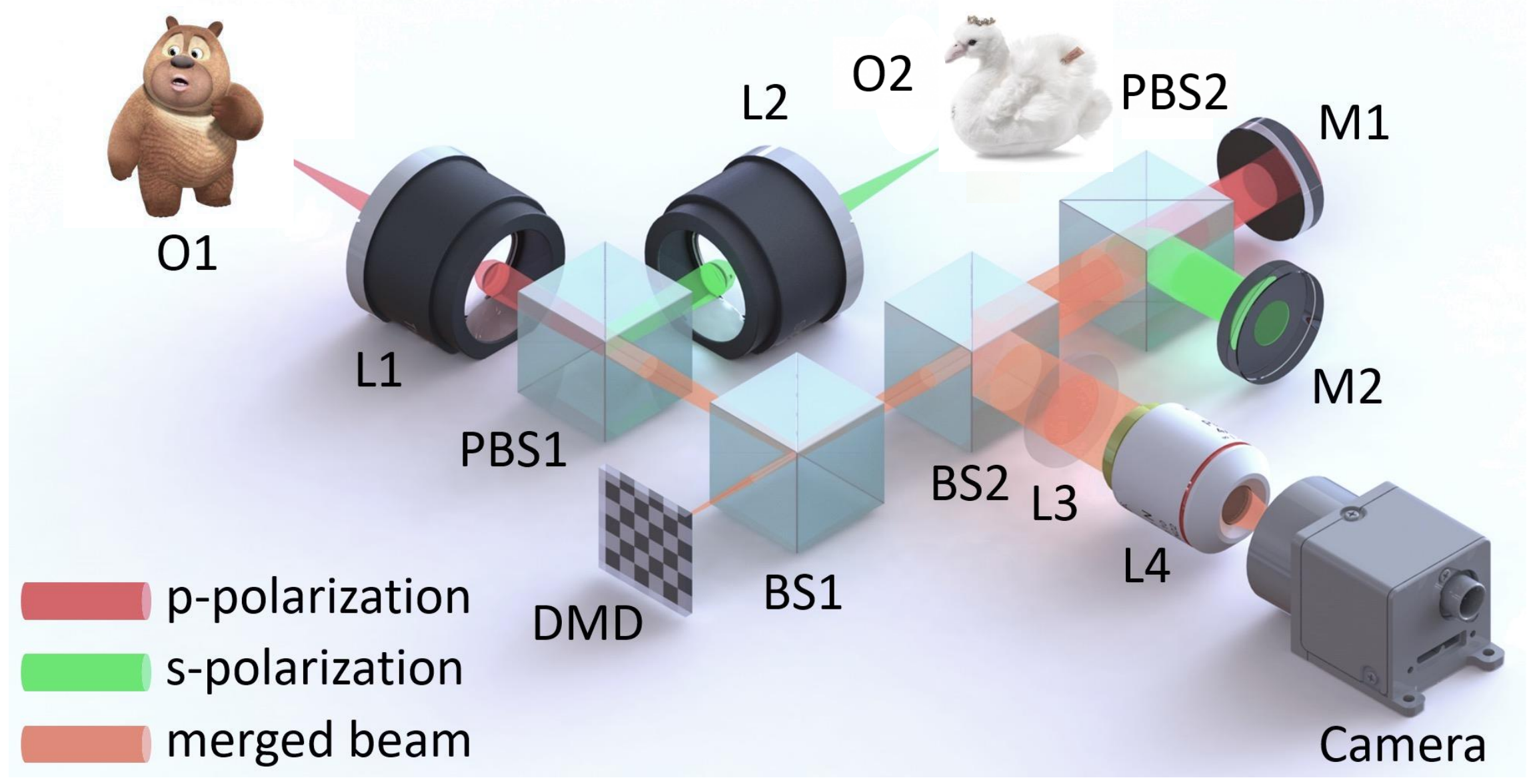}
	\caption{Optical setup of SSTCI~\cite{qiao2020snapshot}. O1, O2: objects from two views; L1, L2, L3, L4: lens; M: mirror; BS: beamsplitter; PBS: polarizing beamsplitter; DMD: digital micromirror device. M1 and M2 have slightly different orientations relative to their incident beams to introduce a lateral displacement (thus imposing different sets of masks on the scenes from two different views) between the modulated images of O1 and O2 on the camera.}
	\label{fig:SSTCI}
\end{figure}
Our SSTCI system was presented in~\cite{qiao2020snapshot}, {which uses} a {\em single camera} to simultaneously capture video streams from two scenes. Between the camera and the scenes, we deploy a single DMD to apply distinguishing spatial modulations (\ie, 2D random-binary masks) to different video frames for both scenes in a manner of element-wise product.
As shown in Fig.~\ref{fig:SSTCI}, the hardware innovation in~\cite{qiao2020snapshot} is to introduce a polarization-dependent transverse displacement of the two views on the detector plane, which is implemented by the polarized beamsplitter (to impose different polarization states onto the scenes from different views) and two mirrors, with different orientations (to shift the modulation patterns).
{It is worth noting that different views can be separated as long as their respective coding patterns are uncorrelated to each other. This condition is met in the SSTCI system in which the shifted versions of the random pattern are mutually uncorrelated as long as the shifting amount is greater than the pattern feature size (size of each random element).} In this case, we have achieved {\em two different sets of masks with each set for the scene of one view, via only using a single DMD and a single camera}. {The system} thus did not scarify the spatial resolution of the DMD and camera and can capture two full views of the scene simultaneously.
A single camera is used to capture both the SCI measurement and side information in~\cite{Yuan17AO}, which can potentially be used to capture two views of the scene, but the solution used therein scarifies the spatial resolution of the camera by half. 

The modulated data streams from the two scenes are then laterally superposed (via beamsplitters) and temporally integrated within one exposure period of the camera, forming a single frame of raw measurement. {In this paper, we aim to propose a reconstruction method to resolve the videos of two scenes from the single measurement.}
{To the best of our knowledge, this is the first deep learning based reconstruction method reported in the literature for joint FoV and temporal CS system and has the potential to extend to multiple views, which will be beneficial to practical robotics and 3D applications (for example, for the left view and right view in the stereo imaging) with a fast inference speed, enjoying the advantages of low-bandwidth, low-memory cost and fast inference.}
\begin{figure*}[!ht]
	\centering
	\includegraphics[width=2\columnwidth]{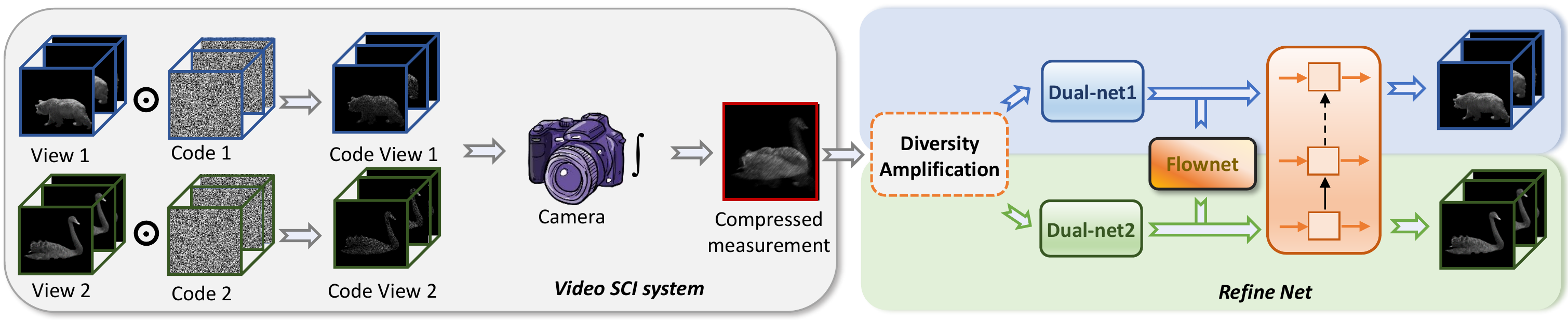}
	\caption{\label{fig:model} {Principle of dual-view video SCI (left) and the proposed network for reconstruction (right). Two FoV dynamic scenes, shown as two branches of images (view 1 and 2) at different timestamps, pass through two sets of dynamic aperture (produced by the same set of masks in Fig.~\ref{fig:SSTCI}), which imposes individual coding patterns. The coded frames from both FoVs are then integrated over time on a  camera, forming a single-frame compressed measurement. This measurement is fed into our proposed model (right) to reconstruct the high-speed video frames of two dynamic scenes.}}
\end{figure*}
\subsection{Mathematical Model of Dual-view SCI}

As depicted in the left part of Fig.~\ref{fig:model} (and also in Fig.~\ref{fig:SSTCI}), in the dual-view video SCI system,  we denote object $O1$ with $B$ temporal channels and $n_x \times n_y$ pixels in the transverse plane as $\Xmat_{1} \in \mathbb{R}^{n_x \times n_y \times B}$ , which {is} modulated by the coding patterns $\Cmat_1 \in {\mathbb R}^{n_x \times n_y \times B}$, correspondingly. Let $\Xmat_{2} \in \mathbb{R}^{n_x \times n_y \times B}$ {denotes} object $O2$, modulated by shifted coding patterns $\Cmat_2 \in {\mathbb R}^{n_x \times n_y \times B}$ (this shifting is introduced by the optical design in Fig.~\ref{fig:SSTCI}). The single-shot measurement $\Ymat \in {\mathbb R}^{n_x \times n_y}$ is given by
\begin{equation}
\label{Eq:YXC}
{ \Ymat = \sum_{b=1}^B \left( \Xmat_1^b \odot  \Cmat_1^b + \Xmat_2^b \odot  \Cmat_2^b \right)+ \Gmat}\,,
\end{equation}
where $\Gmat \in {\mathbb R}^{n_x \times n_y}$ represents the noise and $\odot$ denotes the Hadamard (element-wise) product. The frontal slices $\Xmat_1^b$, $\Cmat_1^b$, $\Xmat_2^b$, $\Cmat_2^b$ with dimension of ${\mathbb R}^{n_x \times n_y}$ denote the $b$-th video frame and the corresponding coding pattern imposed on $O1$ and $O2$, respectively.

Let $\Xmat=\left[\Xmat_{1}, \Xmat_2 \right]_{c3} \in \mathbb{R}^{n_{x} \times n_{y} \times 2B}$
and $\Cmat = \left[\Cmat_1, \Cmat_2\right]_{c3}$ $\in \mathbb{R}^{n_{x} \times n_{y} \times 2B}$, where $[\cdot]_{c3}$ denotes matrix
concatenation operation in the third (temporal) dimension. Then \eqref{Eq:YXC} can be simplified as:
\begin{equation}
\centering
\label{Eq:simple_YXC}
{ \Ymat = \sum_{b=1}^{2B} \Xmat^b \odot \Cmat^b + \Gmat}\,.
\end{equation}
The formulation in \eqref{Eq:simple_YXC} can be expressed by the following vectorized linear equation:
\begin{equation}
\label{Eq:y}
\yv=\Phimat \xv+\gv,
\end{equation}
where $\yv=\operatorname{Vec}(\Ymat) \in \mathbb{R}^{n_{x} n_{y}}$ denotes the vectorized measurement, $\gv=\operatorname{Vec}(\Gmat) \in \mathbb{R}^{n_{x} n_{y}}$ the vectorized noise and $\xv=\operatorname{Vec}(\Xmat) \in \mathbb{R}^{2 n_{x} n_{y} B}$ the desired
signal. Different from traditional compressive sensing \cite{Donoho06ITT}, this sensing
matrix $\Phimat \in \mathbb{R}^{n_{x} n_{y} \times 2 n_{x} n_{y} B}$ in dual-view video SCI is sparse and follows a diagonal structure
\begin{equation}
\Phimat=\left[\operatorname{diag}\left(\operatorname{Vec}\left(\Cmat^{1}\right)\right), \ldots, \operatorname{diag}\left(\operatorname{Vec}\left(\Cmat^{2B}\right)\right)\right].
\end{equation}
Consequently, the compressive sampling rate is equal to  $\frac{1}{2B}$.
Theoretical results have been derived recently in~\cite{Jalali19TIT_SCI} considering this special sensing matrix.
\section{Proposed Framework}
To reconstruct the high-speed frames of two FoVs $\{\Xmat_1^b\}_{b=1}^B$ and $\{\Xmat_2^b\}_{b=1}^B$, we propose an overall architecture of dual video CS reconstruction framework as shown in the right part of Fig.~\ref{fig:model}. The proposed model consists of three components: 1) A {\em diversity amplifier} to distinguish  two scenes. 2) A {\em dual-net separator} to reconstruct each FoV with coarse resolution. 3) An {\em RNN} in conjunction with {\em bidirectional optical flow} to exploit motion information for improving the resolution and refining the reconstructed video frames.
The whole network is trained in an end-to-end manner.

\subsection{Diversity Amplifier}
\begin{figure}[t!]
	\centering
	\includegraphics[width=1.0\columnwidth]{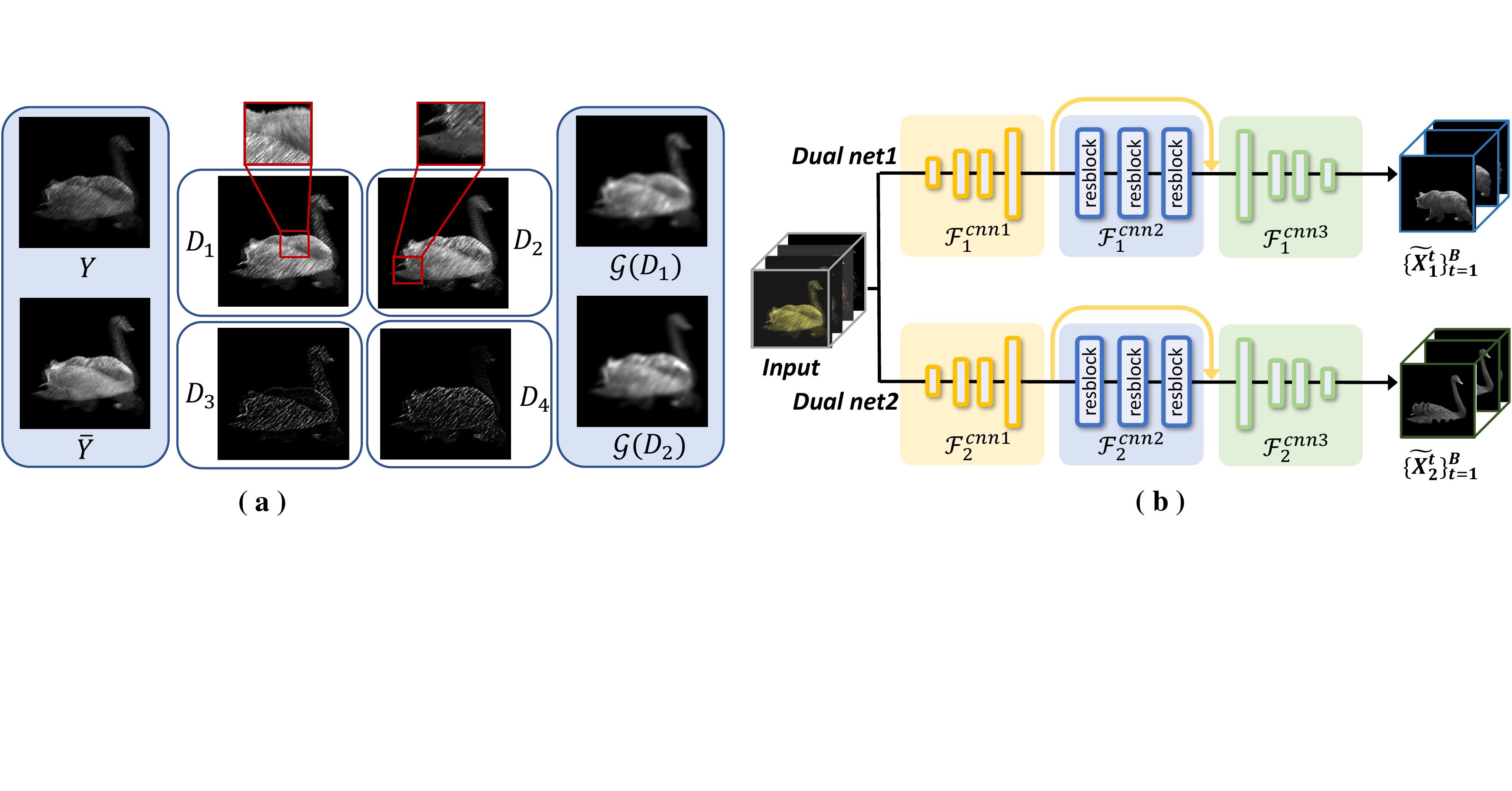}
	\caption{\label{fig:diversity} {Illustration of the original measurement $\Ymat$ (top-left), the normalized measurement $\overline{\Ymat}$ (bottom-left), and four diversity amplified images $\Dmat_1$, $\Dmat_2$, $\Dmat_3$, $\Dmat_4$ (middle) and the smoothed varieties $\mathcal{G}(\Dmat_1),\mathcal{G}(\Dmat_2)$ (right), where we enlarge the details in red boxes for better visualization.}}
\end{figure}

The first step in our proposed network is to distinguish two scenes in the {single} snapshot measurement. Firstly, we normalize the original measurement $\Ymat$ {to balance the integrated energy} by taking all the coding patterns into consideration. Recalling the definition of $\Ymat$ in \eqref{Eq:simple_YXC}, various pixels of $\Ymat$ may gather different numbers of frames from $\{X^b\}_{b=1}^{2B}$ as the integrated energy according to $\{C^b\}_{b=1}^{2B}$, which means some pixels acquire with large energy but some others with low energy. {In other words}, the integrated energy of each pixel is not only related to the value of corresponding position of original scene, but also the corresponding position of the coding patterns. To alleviate the influence of imbalanced energy {and encourage the network to focus on the video content with less disturbs from coding patterns}, we normalize the original measurement {as follows}:

\begin{equation}
\overline{\Ymat} ={\Ymat}\oslash{\left( \sum_{b=1}^{2B} \Cmat^b /(2B)\right)} ,
\end{equation}
{where $\oslash$ denotes the matrix dot (element-wise) division, and the matrix $\sum_{b=1}^{2B} \Cmat^b /(2B)$ refers to the averaged coding patterns with the same size as $\Ymat$, which can be regarded as the corresponding weights of each pixel integrated into the measurement $\Ymat$.
{Fig.~\ref{fig:diversity} shows the illustrations of both the original measurement $\Ymat$ and the normalized measurement $\overline{\Ymat}$ by imposing the energy normalization.} In short, $\overline{\Ymat}$ can be regarded as an approximated averaging image of all the $2B$ high-speed frames $\{\Xmat_1^b\}_{b=1}^B$ and $\{\Xmat_2^b\}_{b=1}^B$, {hoping to normalize the imbalanced energy brought by coding patterns and facilitate the dual-view reconstruction.}
However, the approximated averaging image of all high-speed frames is not sufficient for dual-view SCI, because $\overline{\Ymat}$ averages two scenes without discrimination, expressing the same content of different views.
Thus, to assist the separation of dual views, we develop a diversity amplification preprocessing method by taking different coding patterns of each scene into consideration, aiming to enlarge the differences of two scenes at the first step for better dual-view SCI.}

{As depicted in Fig.~\ref{fig:diversity}, we first construct two precessing methods to explore the dissimilarities between two scenes as follows:} 
\begin{equation}
\begin{array}{l}
\Dmat_1={\Ymat}\oslash{\left( \sum_{b=1}^B \Cmat_1^b /B \right)},\\
\Dmat_2={\Ymat}\oslash{\left( \sum_{b=1}^B \Cmat_2^b /B \right)}.
\end{array}
\end{equation}

{For $\Dmat_1$, each element of $\left( \sum_{b=1}^B \Cmat_1^b/B \right)$ represents the mean coding pattern of the first scene, describing how many corresponding pixels of the first scene are integrated into the measurement $\Ymat$. Here, we use this mean coding pattern to normalize the measurement from the perspective of the first view, which is able to alleviate the influence of the coding patterns of the first view, leaving the non-normalized energy for the second view. As the visualization of $\Dmat_1$ shown in Fig.~\ref{fig:diversity}, the first view (bear) is relatively smoother and visually clearer with less artifacts, \eg, the fur on the bear's back, compared with the second view (swan) shown with obvious noise and artifacts. Similarly, for $\Dmat_2$, the $\left( \sum_{b=1}^B \Cmat_2^b/B \right)$ refers to the corresponding proportion of the second scene integrated into the measurement $\Ymat$, which is utilized to normalize the measurement and results in normalized energy for the second view. Specifically, as $\Dmat_2$ shown in Fig.~\ref{fig:diversity}, the second scene (swan) is much smoother and has less artifacts than the other one (bear), which can be seen from those non-coincident parts such as the neck and tail of the swan. In summary, {element-wise} dividing the measurement by individual mean coding pattern of different scenes, the diversity of different views can be amplified by taking each mean coding pattern into consideration.
Furthermore, in order to sufficiently explore the detected diversity of $\Dmat_1, \Dmat_2$ and obtain the preliminary contours of each view, we develop an elaborated design to further enlarge the differences of dual views, expressed by:
\begin{equation}
\begin{array}{l}
\Dmat_3=\Dmat_1 - {\cal G}(\Dmat_1),\\
\Dmat_4=\Dmat_2 - {\cal G}(\Dmat_2),
\end{array}
\end{equation}
where ${\cal G}$ is utilized to smooth and blur image $\Dmat_1$, \eg, a Gaussian filter, as shown in Fig.~\ref{fig:diversity}. The noise and artifacts of the second scene can be detected by comparing $\Dmat_1$ to its smooth variety ${\cal G}(\Dmat_1)$, denoted by $\Dmat_3$. As we can see from Fig.~\ref{fig:diversity}, by subtracting the smooth variety ${\cal G}(\Dmat_1)$ from $\Dmat_1$, the position in $\Dmat_1$ with non-normalized energy is typically detected expressed as noise and artifacts, resulting in a contour of the swan as the illustration of $\Dmat_3$.
Analogously, $\Dmat_4$ is constructed by detecting the diversity between $\Dmat_2$ and its smoothed version ${\cal G}(\Dmat_2)$, resulting in a contour of the bear as the illustration of $\Dmat_4$. It can be observed that the diversity between different scenes is amplified, meanwhile, capable of preserving the motionless information such as background and motion trails of different scenes, which is beneficial for reconstructing individual scenes for dual-view video SCI.}

\subsection{Dual-net Separator}
Hereby, we construct a dual-net separator (DNS) to generate two videos through two branches, respectively.
In order to make full use of the amplified diversity of different views and
fuse normalize measurement $\Ymat$ in conjunction with various coding patterns, we construct the input of dual-net separator as:
\begin{equation}
\begin{array}{l}
\mathbf{DNS}_1^0=[\overline{\Ymat}, \Dmat_1, \Dmat_2, \Dmat_3, \Dmat_4,
\overline{\Ymat} \odot \Cmat_{1}^1, \overline{\mathbf{Y}} \odot \Cmat_{1}^2, \\
\qquad \qquad \qquad \qquad \qquad \qquad \qquad \ldots, \overline{\mathbf{Y}} \odot \Cmat_{1}^B]_{c3},\\
\mathbf{DNS}_2^0=[\overline{\Ymat}, \Dmat_1, \Dmat_2, \Dmat_3, \Dmat_4,
\overline{\Ymat} \odot \Cmat_{2}^1, \overline{\mathbf{Y}} \odot \Cmat_{2}^2, \\
\qquad \qquad \qquad \qquad \qquad \qquad \qquad \ldots, \overline{\mathbf{Y}} \odot \Cmat_{2}^B]_{c3},\\
\end{array}
\end{equation}
\begin{figure}[t!]
	\centering
	\includegraphics[width=1.0\columnwidth]{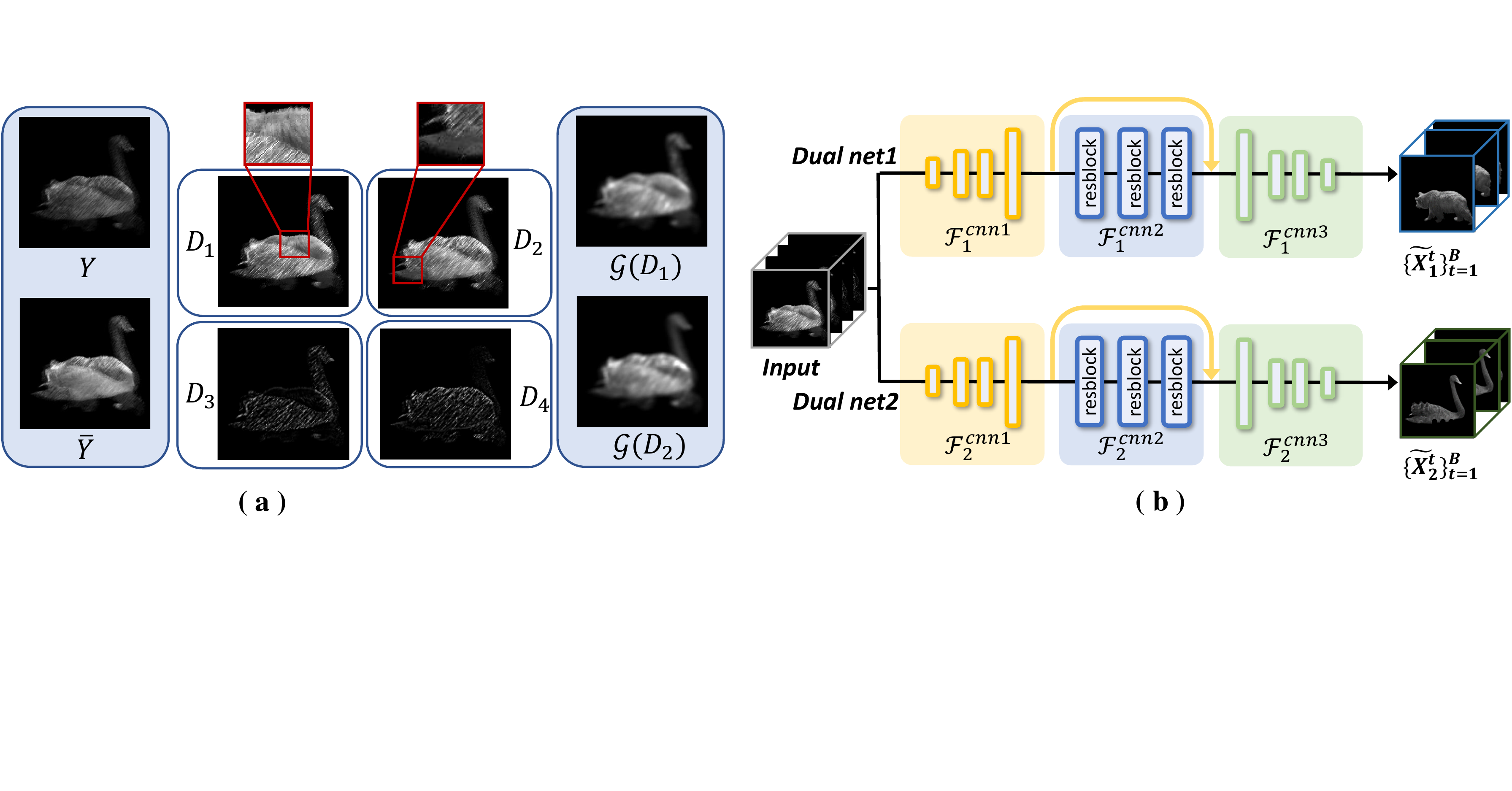}
	\caption{\label{fig:dualnet} Architecture of the dual-net separator.}
\end{figure}
where we approximate the mask-modulated frames of different scenes by element-wise product of normalized measurement $\Ymat$ and the corresponding coding patterns $\{\Cmat_{1/2}^b\}_{b=1}^{B}$ {(hereafter, the subscript $1/2$ means `1 or 2')}. As depicted in Fig.~\ref{fig:dualnet}, the inputs $\mathbf{DNS}_1^0$ and $\mathbf{DNS}_2^0$ are then fed into two branches of convolution networks with shared architecture but different parameters, respectively. Each network branch is composed of three sub-networks, stated as:
\begin{equation}
\begin{split}
&\widetilde{\Xmat}_1=\mathcal{F}_1^{cnn3}([\mathbf{DNS}_1^2,\mathbf{DNS}_1^1]_{c3}),\\
&\mathbf{DNS}_1^2=\mathcal{F}_1^{cnn2}(\mathbf{DNS}_1^1),\\
&\mathbf{DNS}_1^1=\mathcal{F}_1^{cnn1}(\mathbf{DNS}_1^0),\\
&\widetilde{\Xmat}_2=\mathcal{F}_2^{cnn3}([\mathbf{DNS}_2^2,\mathbf{DNS}_2^1]_{c3}),\\
&\mathbf{DNS}_2^2=\mathcal{F}_2^{cnn2}(\mathbf{DNS}_2^1),\\
&\mathbf{DNS}_2^1=\mathcal{F}_2^{cnn1}(\mathbf{DNS}_2^0),\\
\end{split}
\end{equation}
where $\mathcal{F}_{1/2}^{cnn1}$ is used to extract the fine-grained characteristics of input $\mathbf{DNS}_{1/2}^0$, consisting of four convolutional layers. When going deeper, the second subnetwork $\mathcal{F}_{1/2}^{cnn2}$ containing three convolutional resblocks, {is} utilized to explore coarse and global features of the input. The third $\mathcal{F}_{1/2}^{cnn3}$ has a mirror symmetry structure as $\mathcal{F}_{1/2}^{cnn1}$, where fine-to-coarse spatial features of various scales are concatenated together to reconstruct dual videos. In this stage, the two scenes $\widetilde{\Xmat}_1 = \{\widetilde{\Xmat}_1^b\}_{b=1}^B$ and $\widetilde{\Xmat}_2= \{\widetilde{\Xmat}_2^b\}_{b=1}^B$ are separated from each other, and then fed into the refining network as the input providing motion and visual information.

\subsection{Refine Net}
To further explore the spatio-temporal details of videos, the optical flow is employed here to improve the smoothness across time. Moreover, we propagate the optical dynamic information bidirectionally, implemented in conjunction with both a forward and a backward RNN to refine video reconstruction in a sequential manner.
\begin{figure*}[ht!]
\centering
\includegraphics[width=2.0\columnwidth]{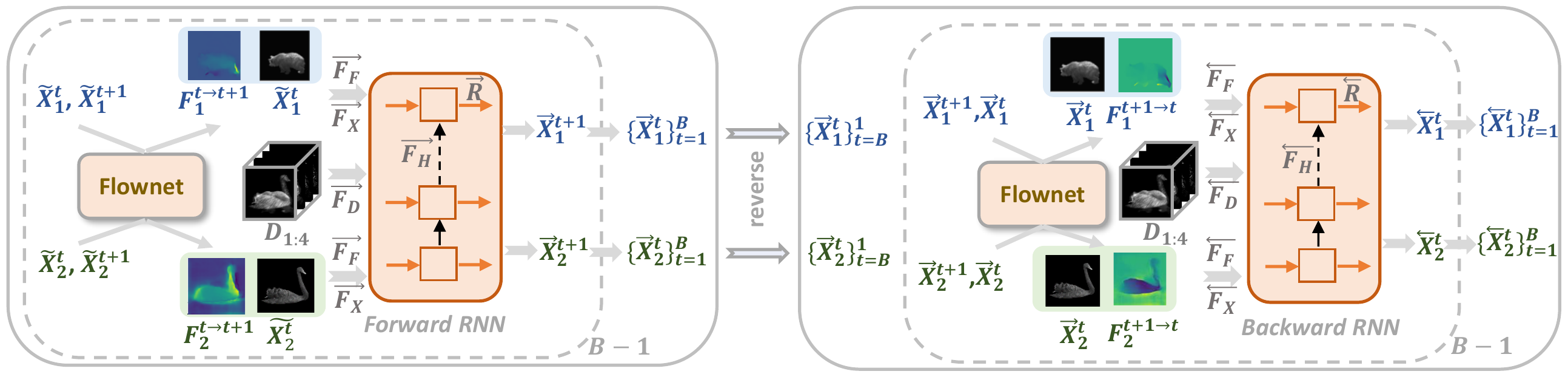}
\caption{\label{fig:birnn} {The architecture of bidirectional recurrent refining network, composed of a forward network (left) recurrently reconstructing each frame forwardly by integrating the corresponding optical flow, and a backward network (right) incorporating with dynamic motion in the reversed order. }}
\end{figure*}
\paragraph{\textbf{Optical Flow Extractor:}}

In order to maintain temporally connected dynamic motions, optical flow is utilized in our work to improve the visual quality {of reconstructed} results. Considering the efficiency and accuracy, we utilize the {\em FlowNet 2.0}~\cite{IlgMSKDB17} as the optical flow extractor, to facilitate video reconstruction by integrating explicit motion information as a guidance.
Our goal is to better explore the dynamic motion between frames bidirectionally. {Taking} the $t$-th and the $(t+1)$-th frame as examples, {the optical flow extraction can be} expressed as:
\begin{equation}
\begin{split}
\Fmat_1^{t\rightarrow t+1}={\rm Flownet}(\widetilde{\Xmat}_1^t, \widetilde{\Xmat}_1^{t+1}),\\
\Fmat_1^{t+1\rightarrow t}={\rm Flownet}(\widetilde{\Xmat}_1^{t+1}, \widetilde{\Xmat}_1^t),\\
\Fmat_2^{t\rightarrow t+1}={\rm Flownet}(\widetilde{\Xmat}_2^t, \widetilde{\Xmat}_2^{t+1}),\\
\Fmat_2^{t+1\rightarrow t}={\rm Flownet}(\widetilde{\Xmat}_2^{t+1}, \widetilde{\Xmat}_2^t),\\
\end{split}
\end{equation}
where $\{\Fmat_1^{t\rightarrow t+1}\}_{t=1}^B$ represents the optical flow from $\widetilde{\Xmat}_1^t$ to $\widetilde{\Xmat}_1^{t+1}$ of the first video scene, while the $\Fmat_1^{t+1\rightarrow t}$ denotes the optical flow {in} the reverse direction. Analogously, $\Fmat_2^{t\rightarrow t+1}$ and $\Fmat_2^{t+1\rightarrow t}$ refer to the bidirectional motions between adjacent frames of the second video scene.

{To make good use of the pre-trained model in~\cite{IlgMSKDB17}, we initialize the weights of Flownet from the released pre-trained model, and jointly fine-tune the parameters under the supervision of the final reconstruction loss.} This is different from previous methods \cite{PerazziKBSS17} which utilize the pre-computed optical flow as an additional input, as our model aims to jointly learn useful motion representations {particularly} for our video SCI task. In the following, we will introduce how to integrate the bidirectional optical flow into the video refining process in a recurrent manner, which {establishes reasonable} connections to RNN for better video reconstruction.

\paragraph{\textbf{Bidirectional Recurrent Refining Network:}}
After extracting the optical flow of dual videos in both the forward and backward order, we integrate the motion information into a bidirectional RNN to refine the reconstruction results of {\em dual-net separator} in a sequential manner, as depicted in Fig.~\ref{fig:birnn}, where the left part corresponds to the forward module and the right part refers to the backward module.

\begin{figure*}[t!]
	\centering
	\includegraphics[width=2\columnwidth]{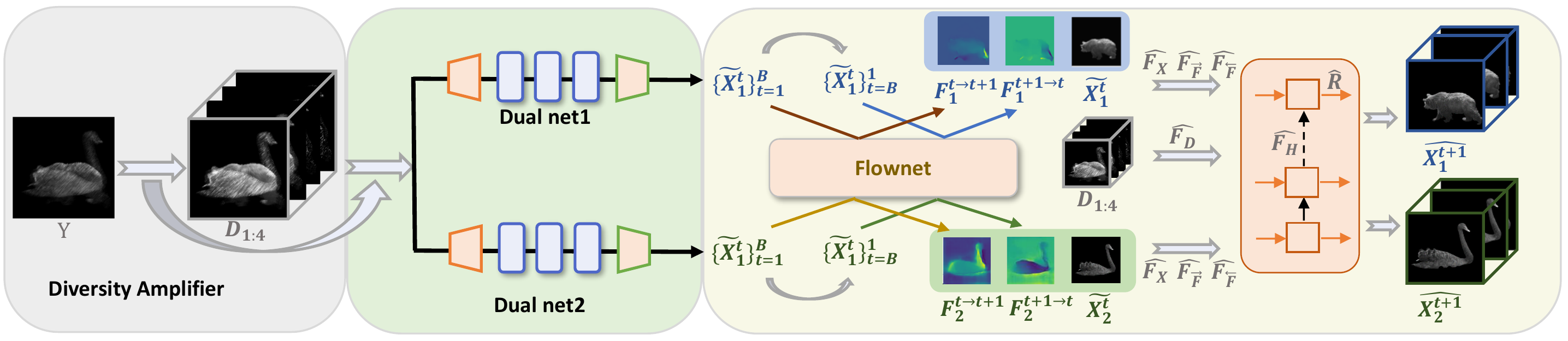}
	\caption{\label{fig:overall} The simplified overall architecture.}
\end{figure*}
\paragraph{Forward Refining Network:} The forward refining network takes the output of DNS $\{\widetilde{\Xmat}_{1/2}^b\}_{b=1}^{B-1}$ as the initial input, and fuses three additional visual information together including the forward motion information, the amplified diversity, and the memory information of RNN. Then each frame of forward refined video $\{\overrightarrow{\Xmat}_{1/2}^b\}_{b=2}^{B}$ is reconstructed recurrently. Taking the $t$-th frame for example, the forward refining process is expressed as:
\begin{equation}
\label{forward_rnn}
\scriptsize
\begin{array}{l}
\overrightarrow{\Xmat}_1^{t+1} = \overrightarrow{\mathcal{R}} \left( \overrightarrow{\Hmat}_1^{t+1} \right),\\
\overrightarrow{\Hmat}_1^{t+1} = \overrightarrow{\mathcal{H}}\left( \left[\overrightarrow{\mathcal{F}_X}(\widetilde{\Xmat}_1^t), \overrightarrow{\mathcal{F}_F}(\Fmat_1^{t\rightarrow t+1}),\overrightarrow{\mathcal{F}_D}(\Dmat_{1: 4}),\overrightarrow{\mathcal{F}_H}(\overrightarrow{\Hmat}_1^t)\right]_{c3} \right),\\
\overrightarrow{\Xmat}_2^{t+1} = \overrightarrow{\mathcal{R}} \left( \overrightarrow{\Hmat}_2^{t+1} \right),\\
\overrightarrow{\Hmat}_2^{t+1} = \overrightarrow{\mathcal{H}}\left( \left[\overrightarrow{\mathcal{F}_X}(\widetilde{\Xmat}_2^t), \overrightarrow{\mathcal{F}_F}(\Fmat_2^{t\rightarrow t+1}),\overrightarrow{\mathcal{F}_D}(\Dmat_{1:4}),\overrightarrow{\mathcal{F}_H}(\overrightarrow{\Hmat}_2^t)\right]_{c3} \right),
\end{array}
\end{equation}
where $\overrightarrow{\mathcal{R}}$ is a CNN block, injecting the hidden representations $\overrightarrow{\Hmat}_1^{t+1}$ to the observation space to output a refined frame $\overrightarrow{\Xmat}_1^{t+1}$, and $\rightarrow$ refers to the forward processing. The hidden representation fuses the visual information of four components together through a CNN resblock denoted by $\overrightarrow{\mathcal{H}}$, including the $t$-th frame $\widetilde{\Xmat}_1^{t+1}$ as a good initialization, the forward optical flow $\Fmat_1^{t\rightarrow t+1}$ providing the motion information, the amplified diversity $\Dmat_{1:4}=[\Dmat_1,\Dmat_2,\Dmat_3,\Dmat_4]_{c3}$ emphasizing the difference between two scenes, and the hidden unit $\Hmat_1^t$ of RNN offering the memory of the previous frames. {In addition}, those visual information are firstly {embedded by} four  parallel CNN blocks, respectively, denoted by $\overrightarrow{\mathcal{F}_X}$, $\overrightarrow{\mathcal{F}_F}$, $\overrightarrow{\mathcal{F}_D}$, $\overrightarrow{\mathcal{F}_H}$, playing the roles of visual feature extractors. Similarly, the video frame $\widetilde{\Xmat}_2^{t}$ of the other scene is refined in the same way. While the forward refining network achieves appealing performance, it neglects the dynamic visual information in the reversed order. 
This encourages us to construct a backward refining network to further improve the reconstruction performance.

\paragraph{Backward Refining Network:} As illustrated in the right part of Fig.~\ref{fig:birnn}, the forwardly refined videos $\{\overrightarrow{\Xmat}_1^{t}\}_{t=1}^B$ and $\{\overrightarrow{\Xmat}_2^{t}\}_{t=1}^B$ are firstly reversed as $\{\overrightarrow{\Xmat}_1^{t}\}_{t=B}^1$ and $\{\overrightarrow{\Xmat}_2^{t}\}_{t=B}^1$, and then taken as the inputs of backward refining network. The structure {of backward refining network} is similar to the forward one, with the main difference on the reversed input and motion information. Due to the opposite order, the $(t+1)$-th frame and the optical flow from $\overrightarrow{\Xmat}_{1/2}^{t+1}$ to $\overrightarrow{\Xmat}_{1/2}^{t}$ are utilized to improve the $t$-th frame, stated as:
\begin{equation}
\label{Backward_rnn}
\small
\begin{array}{l}
\overleftarrow{\Xmat}_1^{t} = \overleftarrow{\mathcal{R}} \left( \overleftarrow{\Hmat}_1^{t} \right),\\
\overleftarrow{\Hmat}_1^{t} = \overleftarrow{\mathcal{H}}\Big( \Big[\overleftarrow{\mathcal{F}_X}(\overrightarrow{\Xmat}_1^{t+1}), \overleftarrow{\mathcal{F}_F}(\Fmat_1^{t+1\rightarrow t}),\overleftarrow{\mathcal{F}_D}(\Dmat_{1:4}), \\
\qquad \qquad \qquad \qquad \qquad \qquad \qquad \quad \overleftarrow{\mathcal{F}_H}(\overleftarrow{\Hmat}_1^{t+1})\Big]_{c3} \Big),\\
\overleftarrow{\Xmat}_2^{t} = \overleftarrow{\mathcal{R}} \left( \overleftarrow{\Hmat}_2^{t} \right),\\
\overleftarrow{\Hmat}_2^{t} = \overleftarrow{\mathcal{H}}\Big( \Big[\overleftarrow{\mathcal{F}_X}(\overrightarrow{\Xmat}_2^{t+1}), \overleftarrow{\mathcal{F}_F}(\Fmat_2^{t+1\rightarrow t}),\overleftarrow{\mathcal{F}_D}(D_{1:4}),\\
\qquad \qquad \qquad \qquad \qquad \qquad \qquad \quad \overleftarrow{\mathcal{F}_H}(\overleftarrow{\Hmat}_2^{t+1})\Big]_{c3} \Big),
\end{array}
\end{equation}

where $\leftarrow$ refers to the backward processing, and all the feature extractors have exactly the same architectures
corresponding to the forward refining network, but without sharing parameters. As a result, the reconstructed videos of two scenes are improved by integrating the opposite dynamic information, which measures the spatio-temporal shifts of the intensity in a reversed order. {{Overall, the bidirectional designation} has three benefits: 1) the forwardly refined results are more accurate and able to provide more detailed motion and visual information; 2) the backward optical flow is informative to provide the reversed motion for the predicting frame; 3) the bidirectional mechanism provides more gradient propagation paths when training with the gradient descent algorithm, which helps to jointly optimize the forward and backward reconstruction.}

\subsection{Simplified Overall Architecture}
Although the architecture composed of both forward and backward refining network improves reconstruction results bidirectionally, the drawback is its large computational cost and huge memory requirement.
To address this challenge, and aiming to save half of the required memory and {take the advantages of }the bidirectional motion information, we further condense the forward and backward refining network together, as shown in Fig.~\ref{fig:overall}. Firstly, the outputs of DNS $\{\widetilde{\Xmat}_{1/2}^{t}\}_{t=1}^B$ are fed into the flownet to extract the optical flow $\{\Fmat_1^{t \rightarrow t+1}\}_{t=1}^{B-1}$ in the forward order. Similarly, the reversed $\{\widetilde{\Xmat}_{1/2}^{t}\}_{t=B}^1$ are fed into the flownet to obtain the optical flow $\{\Fmat_1^{t+1 \rightarrow t}\}_{t=B-1}^{1}$ in the backward order. Then the bidirectional motion information is taken as the input of RNN, stated as:
\begin{equation}
\label{bidirection_rnn}
\small
\begin{array}{l}
\widehat{\Xmat}_1^{t+1} = \widehat{\mathcal{R}} \left( \widehat{\Hmat}_1^{t+1} \right), \qquad \widehat{\Xmat}_2^{t+1} = \widehat{\mathcal{R}} \left(\widehat{\Hmat}_2^{t+1} \right),\\
\widehat{\Hmat}_1^{t+1} = \widehat{\mathcal{H}}\Big( \Big[\widehat{\mathcal{F}_X}(\widetilde{\Xmat_1^{t}}),
\widehat{\mathcal{F}_{\overrightarrow{F}}}(\Fmat_1^{t\rightarrow t+1}), \widehat{\mathcal{F}_{\overleftarrow{F}}}(\Fmat_1^{t+1\rightarrow t}) ,\\
\qquad \qquad \qquad \qquad \qquad \qquad \widehat{\mathcal{F}_D}(\Dmat_{1:4}),\widehat{\mathcal{F}_H}(\widehat{\Hmat}_1^{t})\Big]_{c3} \Big),\\
\widehat{\Hmat}_2^{t+1} = \widehat{\mathcal{H}}\Big( \Big[\widehat{\mathcal{F}_X}(\widetilde{\Xmat_2^{t}}),
\widehat{\mathcal{F}_{\overrightarrow{F}}}(\Fmat_2^{t\rightarrow t+1}), \widehat{\mathcal{F}_{\overleftarrow{F}}}(\Fmat_2^{t+1\rightarrow t}),\\
\qquad \qquad \qquad \qquad \qquad \qquad \widehat{\mathcal{F}_D}(\Dmat_{1:4}), \widehat{\mathcal{F}_H}(\widehat{\Hmat}_2^{t})\Big]_{c3} \Big),
\end{array}
\end{equation}
where the main modification is that the two CNN blocks $\widehat{\mathcal{F}_{\overrightarrow{\Fmat}}}$ and $\widehat{\mathcal{F}_{\overleftarrow{\Fmat}}}$ are simultaneously utilized to inject bidirectional optical flow into the hidden space, which facilitates reconstruction of the $(t+1)$-th frame by predicting from the dynamic information $\Fmat_{1/2}^{t\rightarrow t+1}$ and the backward reasoning from $\Fmat_{1/2}^{t+1\rightarrow t}$. All the feature extractors $ \{ \widehat{\mathcal{R}},\widehat{\mathcal{H}},\widehat{\mathcal{F}_X},\widehat{\mathcal{F}_{\overrightarrow{F}}},\widehat{\widehat{\mathcal{F}_{\overleftarrow{F}}}},\widehat{\mathcal{F}_D},\widehat{\mathcal{F}_H} \}$ have the same architectures as introduced in the forward and backward refining network. We find that this framework is able to save half the memory with little performance degradation, taking the advantages of both forward and backward optical flows.

To sum up, the whole network is composed of the dual-net separator and the simplified bidirectional recurrent refining network, as shown in Fig.~\ref{fig:overall}. To minimize the reconstruction error of all the frames of dual view, we employ the mean square error (MSE) as the loss function {for training} the end-to-end deep network, stated as:
\begin{equation}
\begin{array}{l}
\mathcal{L} =\alpha \widetilde{\mathcal{L}}+\widehat{\mathcal{L}}, \\
\widetilde{\mathcal{L}} =\sum_{t=1}^{B}\left\|\widetilde{\mathbf{X}}_1^{t}-\mathbf{X}_1^{t}\right\|_{2}^{2} + \sum_{t=1}^{B}\left\|\widetilde{\mathbf{X}}_2^{t}-\mathbf{X}_2^{t}\right\|_{2}^{2}, \\
\widehat{\mathcal{L}}=\sum_{t=1}^{B}\left\|\widehat{\mathbf{X}}_1^{t}-\mathbf{X}_1^{t}\right\|_{2}^{2}+\sum_{t=1}^{B}\left\|\widehat{\mathbf{X}}_2^{t}-\mathbf{X}_2^{t}\right\|_{2}^{2},\\
\end{array}
\end{equation}
where $\widetilde{\mathcal{L}}$ and $\widehat{\mathcal{L}}$ represent the MSE loss of dual-net separator and refine net, respectively, and $\alpha$ is a trade-off parameter, which is set to 1 in our experiments. During testing, we can achieve high-quality reconstructed videos in a short time with the well-learned network parameters.

\section{Experiments}
{In this section, we compare the proposed optical flow-aided recurrent neural network (OFaNet) with other methods on both simulation and real datasets for dual-view SCI, and further extend it to single-view SCI for a broader scope.}
\subsection{Datasets, Training and Counterparts}
\paragraph{Datasets:}
We first evaluate the proposed OFaNet on our collected six simulation data including \texttt{Bear\&Blackswan}, \texttt{Boy\&Girl}, \texttt{Running Cars}, \texttt{Bike\&Bus}, \texttt{Cow\&Dog} and \texttt{Hike} \texttt{\&Hockey}, respectively, where $B = 10$ video frames of two individual $256\times256$ scenes (i.e. totally 20 frames) are compressed into a single measurement for each dataset. We also demonstrate experiments on simulation data with different compressive rates ($B = \{ 6, 14\}$). To evaluate the effectiveness of the proposed model in a broader scope, we also demonstrate experiments on the {\em single-view video SCI} on six widely used grayscale simulation datasets including \texttt{Kobe}, \texttt{Runner}, \texttt{Drop}, \texttt{Traffic}, \texttt{Aerial} and \texttt{Vehicle}, where 8 video frames are compressed into a single measurement.
{Furthermore, we also evaluate OFaNet on two real dual-view datasets with $B = 10$, one real dual-view dataset with $B = 20$ \cite{qiao2020snapshot}, captured by the real SSTIC camera, and three real single-view datasets.}
\paragraph{Implementation Details:}
For simulation, We randomly crop patch cubes $256 \times256 \times B$ from original scenes in DAVIS2017 \cite{pont20172017} and randomly select two video sequences at each time to be modulated with different patterns, and obtain the training dataset containing about 32,000 data pairs. We set the maximum training epoch as 90, the batch size as 2, and start training with the initial learning rate of $3 \times 10^{-4}$, which is decreased by 10\% every 10 epochs. Our network is implemented in Pytorch and trained with the Adam optimizer \cite{adam}. The experiments are conducted on a NVIDIA RTX 8000 GPU, and it takes about 3 days to complete the training of the entire network. {For the real SSTIC dataset with the size of 650 $\times$ 650 $\times$ 10 ($B=10$), we randomly crop patch cubes (650 $\times$ 650 $\times$ 10) from the same training dataset DAVIS2017, and obtain about 12000 data pairs for training with the initial learning rate of $2 \times 10^{-5}$. Similarly, we generate 12000 data pairs with the size of 650 $\times$ 650 $\times$ 20 for training another real dataset with $B=20$.} The detailed architecture of OFaNet is presented in Table~\ref{setting}.
\begin{table}[h!]\huge
\centering
\caption{Network architecture of the proposed OFaNet.}
\resizebox{0.5\textwidth}{!}{
\begin{tabular}{c|c|c|c|c}
\toprule[1pt]
    \textbf{ Module } & \textbf{ Operation } & \textbf{ Kernel } & \textbf{ Stride }  & \textbf{ Output Size }
\\ \hline
\multirow{4}{*}{$\mathcal{F}_{1/2}^{cnn1}$}
& conv. & $5\times 5$ &1 & $256\times256\times32$\\
& conv. & $3\times 3$ &1 & $256\times256\times64$\\
& conv. & $1\times 1$ &1 & $256\times256\times64$\\
& conv. & $3\times 3$ &2 & $128\times128\times64$\\ \hline
\multirow{7}{*}{$\mathcal{F}_{1/2}^{cnn2}$}
& $cnn-resblock$ & $\begin{bmatrix}
    3\times 3\\
    1\times 1\\
    3\times 3\\
\end{bmatrix}\times 3$ & 1 & $128\times128\times 64$\\ \cline{2-5}
& $cnn-resblock$ & $\begin{bmatrix}
    3\times 3\\
    1\times 1\\
    3\times 3\\
\end{bmatrix}\times 3$ & 1 & $128\times128\times 64$\\ \cline{2-5}
& $cnn-resblock$ & $\begin{bmatrix}
    3\times 3\\
    1\times 1\\
    3\times 3\\
\end{bmatrix}\times 3$ & 1 & $128\times128\times 64$\\
\hline
\multirow{4}{*}{$\mathcal{F}_{1/2}^{cnn3}$}
& deconv. & $3\times 3$ &2 & $256\times256\times64$\\
& conv. & $1\times 1$ &1 & $256\times256\times64$\\
& conv. & $3\times 3$ &1 & $256\times256\times32$\\
& conv. & $1\times 1$ &1 & $256\times256\times20$\\ \hline
\multirow{5}{*}{$\mathcal{F}_X$}
& conv. & $5\times 5$ &1 & $256\times256\times 20$\\
& conv. & $1\times 1$ &1 & $256\times256\times 20$\\
& conv. & $3\times 3$ &1 & $256\times256\times 20$\\
& conv. & $1\times 1$ &1 & $256\times256\times 20$\\
& conv. & $3\times 3$ &1 & $256\times256\times 20$\\
 \hline
\multirow{3}{*}{$\mathcal{F}_{\overrightarrow{F}}$}
& conv. & $5\times 5$ &1 & $256\times256\times 40$\\
& conv. & $3\times 3$ &1 & $256\times256\times 40$\\
& conv. & $1\times 1$ &1 & $256\times256\times 40$\\
 \hline
\multirow{3}{*}{$\mathcal{F}_{\overleftarrow{F}}$}
& conv. & $5\times 5$ &1 & $256\times256\times 40$\\
& conv. & $3\times 3$ &1 & $256\times256\times 40$\\
& conv. & $1\times 1$ &1 & $256\times256\times 40$\\ \hline
\multirow{5}{*}{$\mathcal{F}_D$}
& conv. & $5\times 5$ &1 & $256\times256\times 20$\\
& conv. & $1\times 1$ &1 & $256\times256\times 20$\\
& conv. & $3\times 3$ &1 & $256\times256\times 20$\\
& conv. & $1\times 1$ &1 & $256\times256\times 20$\\
& conv. & $3\times 3$ &1 & $256\times256\times 20$\\ \hline
\multirow{9}{*}{$\mathcal{F}_H$}
& $cnn-resblock$ & $\begin{bmatrix}
    3\times 3\\
    1\times 1\\
    3\times 3\\
\end{bmatrix}\times 3$ & 1 & $256\times256\times 40$\\ \cline{2-5}
& $cnn-resblock$ & $\begin{bmatrix}
    3\times 3\\
    1\times 1\\
    3\times 3\\
\end{bmatrix}\times 3$ & 1 & $256\times256\times 40$\\ \cline{2-5}
& conv. & $1\times 3$ &1 & $256\times256\times 20$\\
& conv. & $3\times 1$ &1 & $256\times256\times 20$\\
& conv. & $1\times 3$ &1 & $256\times256\times 10$\\ \hline
\multirow{6}{*}{$\mathcal{R}$}
& conv. & $3\times 3$ &1 & $256\times256\times 40$\\
& conv. & $1\times 1$ &1 & $256\times256\times 30$\\
& conv. & $3\times 3$ &1 & $256\times256\times 20$\\
& conv. & $1\times 1$ &1 & $256\times256\times 20$\\
& conv. & $3\times 3$ &1 & $256\times256\times 20$\\
& conv. & $1\times 1$ &1 & $256\times256\times 1$\\ \hline
\hline
\bottomrule
\end{tabular}}
\label{setting}
\end{table}

\paragraph{Counterpart Methods and Performance Metrics:}
As mentioned before, the PnP-TV-FFD  algorithm~\cite{qiao2020snapshot} has been proposed for dual-view SCI reconstruction.  {Further more, GAP-TV \cite{Yuan16ICIP_GAP} is a widely used efficient baseline, which is able to provide decent results within short time. The previous state-of-the-art optimization based algorithm DeSCI~\cite{Liu18TPAMI} is also employed to dual-view video SCI as a strong baseline, however, which always suffers from slow reconstruction speed. For comparison with the deep learning based methods, we employ the U-net~\cite{Qiao2020_APLP,Unet_RFB15a}, ADMMnet~\cite{Ma19ICCV} and state-of-the-art network BIRNAT~\cite{Cheng20ECCV_BIRNAT} to the dual-view SCI tasks for comparison. In the following, we compare our proposed method against these six methods.}
\begin{itemize}
\item{\textbf{GAP-TV}} \cite{Yuan16ICIP_GAP}: This algorithm models the inverse problem of video SCI as a total variation minimization problem. It aims to solve
\begin{equation}
 \widehat{\xv}=\argmin_{\xv}\|\mathrm{TV}(\xv)\|, \quad \text { s.t. }~~ \yv=\boldsymbol{\Phi} \xv,
\end{equation}
where $\mathrm{TV}(x)$ indicates the total variation of the signal, imposing the sparsity on the gradient of signal. GAP-TV employs 100 iterations for the dual-view SCI video reconstruction.
\item{\textbf{PnP-TV-FFD}} \cite{qiao2020snapshot}: This algorithm utilizes the deep denoising priors in the plug-and-play framework for efficient reconstruction, which solves the following problem
\begin{equation}
\widehat{\xv}=\argmin_{\xv} \frac{1}{2}\|\yv-\Phimat \xv\|_{2}^{2}+\lambda \mathrm{TV}(\xv)+\rho \Thetamat(\xv),
\end{equation}
where $\Thetamat(\xv)$ denotes the deep denoising prior, \ie, the fast and flexible denoising network (FFDNet)~\cite{Zhang18TIP_FFDNet}. PnP-TV-FFD uses 500 iterations for the best performance.
{\item{\textbf{DeSCI}~\cite{Liu18TPAMI}}: This SCI reconstruction algorithm boosted the reconstruction quality of single-view SCI, which imposes a weighted nuclear norm on the mathematical model of SCI system. DeSCI integrates the nonlocal self-similarity of videos and the rank minimization with the SCI decoding problem. By applying the nuclear norm of nonlocal similar patch-groups in the video frames, the signals can be recovered with the minimized rank under the alternating direction method of multipliers (ADMM) framework. The iteration number Max-Iter is fixed to 60 as suggested in~\cite{Liu18TPAMI}.}
\item{\textbf{U-net}}~\cite{Qiao2020_APLP}: {This deep learning based network utilizes a specially designed CNN architecture to capture the local structures for reconstructing video in an end-to-end manner.} For fair comparison, we use the same training datasets as utilized in our proposed model, and employ the same hyperparameters as set in the original paper. We train 80 epochs for U-net on a NVIDIA RTX 8000 GPU, and it takes about 1.5 days to complete the training of the entire network.
{\item{\textbf{ADMM-net}~\cite{Ma19ICCV}}: A deep tensor ADMM-Net provides high-quality decoding for video SCI systems in seconds. This network unfolds inference iterations of a standard tensor ADMM algorithm into a layer-wise structure based on deep neural network, which learns the domain of low-rank tensor through computationally efficient training. We train 50 epochs until the performance converges.}
 \item{\textbf{BIRNAT}~\cite{Cheng20ECCV_BIRNAT}}: This method firstly employs the recurrent networks to SCI and achieves state-of-the-art performance for single-view SCI reconstruction. BIRNAT employs a deep CNN with a self-attention map to reconstruct the first frame, then a forward RNN and a backward RNN to generate the following frames in a sequential manner. BIRNAT also utilizes the adversarial training for performance improvement. We train BIRNAT on an NVIDIA RTX 8000 GPU with the learning rate $1 \times 10^{-4}$ till the performance is convergent, which takes about 12 days for training due to the doubled compression rate corresponding to dual-view SCI.
\end{itemize}

{We evaluate the above algorithms and our proposed OFaNet on six simulation datasets and three real datasets captured by the SSTCI system (dual-view SCI). Furthermore, we also employ the proposed OFaNet and the comparison methods to the single-view SCI system, evaluating the performances on six benchmark simulation datasets and three real datasets captured by real single-view video SCI cameras \cite{Patrick13OE,Qiao2020_APLP}, to validate the effectiveness of OFaNet in a broader scope.}

\begin{table*}[htbp!]
\caption{{The average results of PSNR in dB (left entry in each item) and SSIM (right entry in each item) and running time per measurement/shot in seconds by different algorithms on dual-view simulation datasets.}}
\label{table_simulation}
\centering
\resizebox{\textwidth}{!}{
\begin{tabular}{c|c|c|c|c|c|c|c}
\hline \hline
Dataset &  GAP-TV~\cite{Yuan16ICIP_GAP}  &  PnP-TV-FFD~\cite{qiao2020snapshot}  & DeSCI~\cite{Liu18TPAMI} & U-net~\cite{Qiao2020_APLP}   & ADMM-net~\cite{Ma19ICCV}  & BIRNAT~\cite{Cheng20ECCV_BIRNAT}  &  OFaNet   \\ \hline
Bear &22.82 \quad 0.58 &23.01 \quad 0.58 &24.33 \quad 0.60 &23.70 \quad 0.56 &24.91 \quad 0.60  &24.61 \quad 0.60 &\textbf{25.40} \quad \textbf{0.64} \\
Blackswan &22.50 \quad 0.58 &23.71 \quad 0.58 &24.10 \quad 0.58 &24.11 \quad 0.53 & 24.69 \quad 0.57 &24.42 \quad 0.56 &\textbf{25.76} \quad \textbf{0.65} \\ \hline
Boy &23.75 \quad 0.56 &24.00 \quad 0.56 &25.58 \quad 0.56 &24.48 \quad 0.51 &25.79 \quad 0.54 &25.69 \quad 0.54 &\textbf{26.12} \quad \textbf{0.57} \\
Girl &22.07 \quad 0.61 &22.28 \quad 0.61 &24.41 \quad 0.64 &22.61 \quad 0.53 &24.02 \quad 0.60 &23.96 \quad 0.60 &\textbf{25.70} \quad \textbf{0.68} \\ \hline
Car-b &20.75 \quad 0.61 &21.11 \quad 0.62 &21.93 \quad 0.65 &21.06 \quad 0.57 & 22.08 \quad 0.63 &22.00 \quad 0.64 &\textbf{23.15} \quad \textbf{0.69} \\
Car-w &21.31 \quad 0.64 &21.65 \quad 0.64 &22.74 \quad 0.72 &21.63 \quad 0.59 & 22.41 \quad 0.65 &22.23 \quad 0.66 &\textbf{23.73} \quad \textbf{0.74} \\ \hline
Bike &24.38 \quad 0.67 &22.00 \quad 0.56 &26.65 \quad 0.74 &23.47 \quad 0.57 & 25.23 \quad 0.64 &25.06 \quad 0.67 &\textbf{26.30} \quad \textbf{0.71}\\
Bus &22.31 \quad 0.72 &21.86 \quad 0.66 &28.20 \quad 0.83 &22.25 \quad 0.65 & 23.72 \quad 0.71 &23.55 \quad 0.73 &\textbf{25.43} \quad \textbf{0.79}\\ \hline
Cow &21.63 \quad 0.52 &21.49 \quad 0.48 &22.45 \quad 0.53 &21.91 \quad 0.47 & 22.94 \quad 0.53 &22.65 \quad 0.53 &\textbf{23.15} \quad \textbf{0.55}\\
Dog &23.69 \quad 0.58 &23.93 \quad 0.51 &24.43 \quad 0.53 &24.55 \quad 0.51 & 25.23 \quad 0.54 &24.88 \quad 0.54 &\textbf{25.94} \quad \textbf{0.60}\\ \hline
Hike &21.90 \quad 0.51 &21.48 \quad 0.47 &23.28 \quad 0.57 &22.12 \quad 0.45 & 23.46 \quad 0.53 &23.15 \quad 0.53 &\textbf{23.85} \quad \textbf{0.57}\\
Hockey &22.75 \quad 0.67 &22.89 \quad 0.61 &26.04 \quad 0.78 &23.72 \quad 0.62 & 25.14 \quad 0.69 &24.93 \quad 0.72 &\textbf{26.85} \quad \textbf{0.79}\\ \hline\hline
Average &22.49 \quad 0.60 &22.54 \quad 0.57 &24.51 \quad 0.64 &22.97 \quad 0.55 & 24.14 \quad 0.60 &23.93 \quad 0.61 &\textbf{25.12} \quad \textbf{0.67} \\
Time(s) &16.9 &52.3 &21600.4 &\textbf{0.0216} &0.0629 &0.5132 &0.4893\\ \hline \hline
\end{tabular}}
\end{table*}

Both peak-signal-to-noise ratio (PSNR) and structural similarity (SSIM)~\cite{Wang04imagequality} are used as metrics for assessing the performance on simulation datasets. PSNR is the ratio between the maximum power and the
power of residual errors from the "reference", while SSIM quantifies the visual quality by considering the structural similarity, {which is more consistent with the human visual perception.
Besides, we measure the running time of video reconstruction at the testing stage to evaluate the applicability in real-time applications.}

\subsection{Simulation Data}

We first test the algorithms on six dual-view simulated data: \texttt{Bear\&Blackswan}, \texttt{Boy\&Girl}, \texttt{Running Cars},  \texttt{Bike\&Bus},  \texttt{Cow\&Dog} and  \texttt{Hike\&Hockey}. {The performance comparisons on the six benchmark datasets are summarized in Table \ref{table_simulation}, using different algorithms, \ie, GAP-TV, PnP-TV-FFD, DeSCI, U-net, ADMMnet, BIRNAT and OFaNet. The corresponding visualization results of selected reconstructed frames are shown in Fig.~\ref{fig:result_simulation} with full reconstructed videos shown in the supplementary material (SM).} It can be observed that:

\begin{figure*}[htbp!]
	\centering
	\includegraphics[width=2.0\columnwidth]{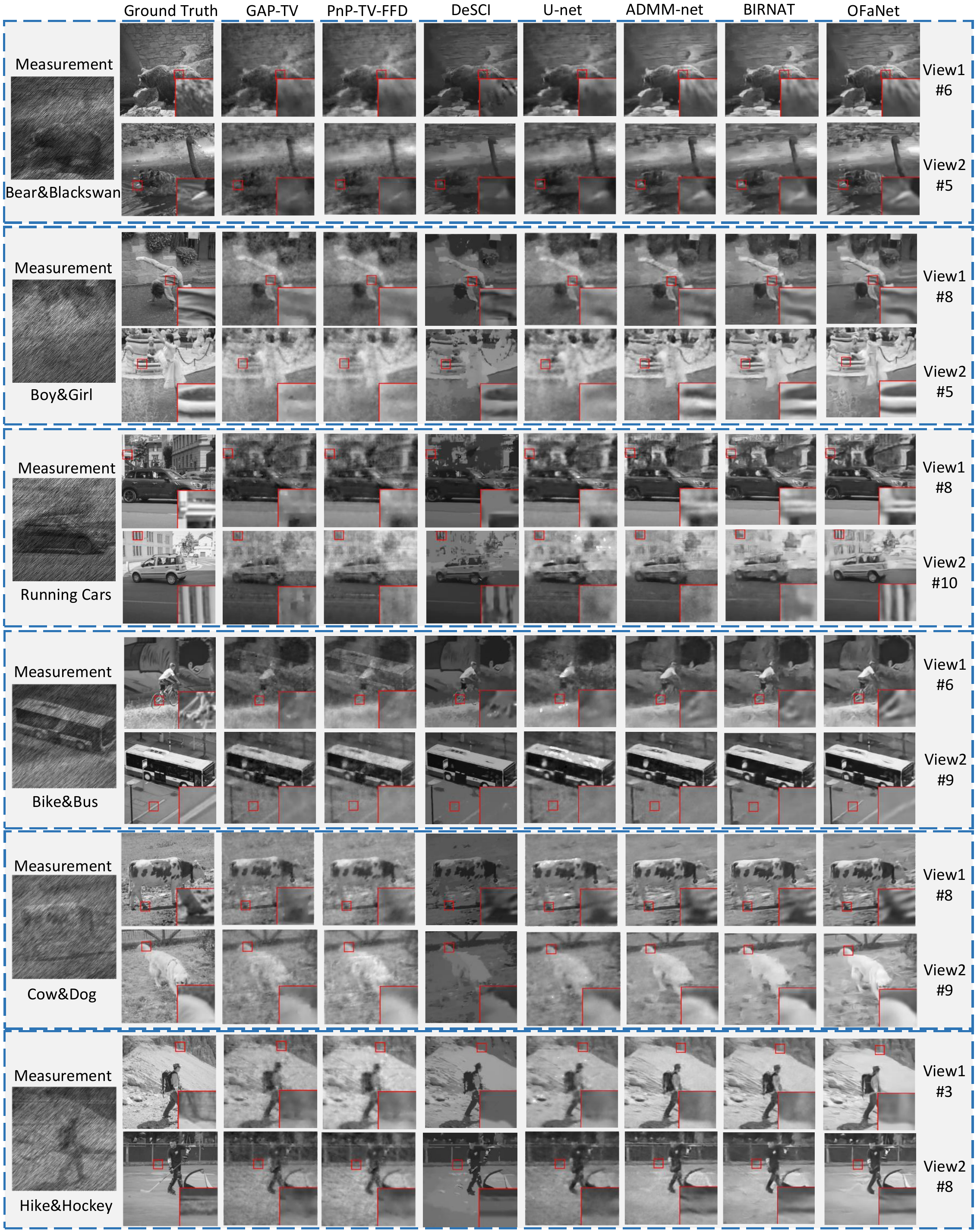}
	\caption{ Reconstructed results of different algorithms on six simulated dual-view SCI datasets. Left: measurements, right: selected reconstructed frames from View 1 and View 2 using different algorithms.}
	\vspace{-3mm}
		\label{fig:result_simulation}
\end{figure*}

{$i$) OFaNet leads to the best performance compared to other algorithms on all the six datasets. In specific, the average gains of OFaNet over \{GAP-TV, PnP-TV-FFD, DeSCI, U-net, ADMM-net and BIRNAT\} are as much as \{2.63, 2.58, 0.61, 2.15, 0.98, 1.19\}dB on PSNR and \{0.07, 0.10, 0.03, 0.12, 0.07, 0.05\} on SSIM. Intuitively, OFaNet provides superior performance owing to the motion smoothness provided by the optical flow and the elaborately designed diversity amplifier which better fits for the target task.}

{$ii$) Selected reconstructed frames are shown in Fig.~\ref{fig:result_simulation}. It can be observed that severe structured artifacts and noise are shown in the reconstructed images of GAP-TV and PnP-TV-FFD caused by the incomplete separation of dual scenes from the single measurement. For example, the vague shape of the bus can be seen from the reconstructed bike, which leads to severe image distortion. DeSCI smooths out some details in the reconstructed video, and results in some artifacts making the reconstructed video more visually 'unrealistic'. Unet results in some white spots randomly located in the reconstructed images, and the edges and corners are blurry. Although the BIRNAT has lower PSNR than ADMM-net, the reconstructed images of BIRNAT are visually cleaner than ADMM-net, corresponding to higher SSIM and contributing to its powerful representative ability. {In general}, the proposed OFaNet provides finer details and clearer contours, owing to the amplified diversity and good representative ability of the DNS and refine net.}

{$iii$) Interestingly, compared with GAP-TV, the methods PnP-TV-FFD and U-net both obtain higher PSNR but lower SSIM, which result in the image distortions in Fig.~\ref{fig:result_simulation} as these so-called natural scenes look 'unnatural'. Similarly, the ADMM-net also results in higher PSNR but lower SSIM than BIRNAT, corresponding to a blurred visualization. It can be seen that the proposed OFaNet gains improvements both on PSNR and SSIM and provides cleaner and sharper reconstruction corners, which might contribute to the fine-to-cross structure and the sequential dependencies constructed by RNN.}

{$iv$) Although the deep learning based methods (U-net, ADMM-net, BIRNAT and OFaNet) require more time for training, these methods are able to reconstruct videos within 1 second at the testing stage, which are significantly faster than the iterative optimization based algorithms. It is worth noting that our proposed OFaNet achieves more than 40,000 times faster than DeSCI ( the runner-up on PSNR and SSIM) at the testing stage. Although U-net is the fastest method, it suffers from poor reconstruction quality as shown in Fig.~\ref{fig:result_simulation}. Thus OFaNet is a pretty good compromise considering the trade-off between {reconstruction} quality and time.}

\begin{figure}[th!]
	\centering
	\includegraphics[width=1.0\columnwidth]{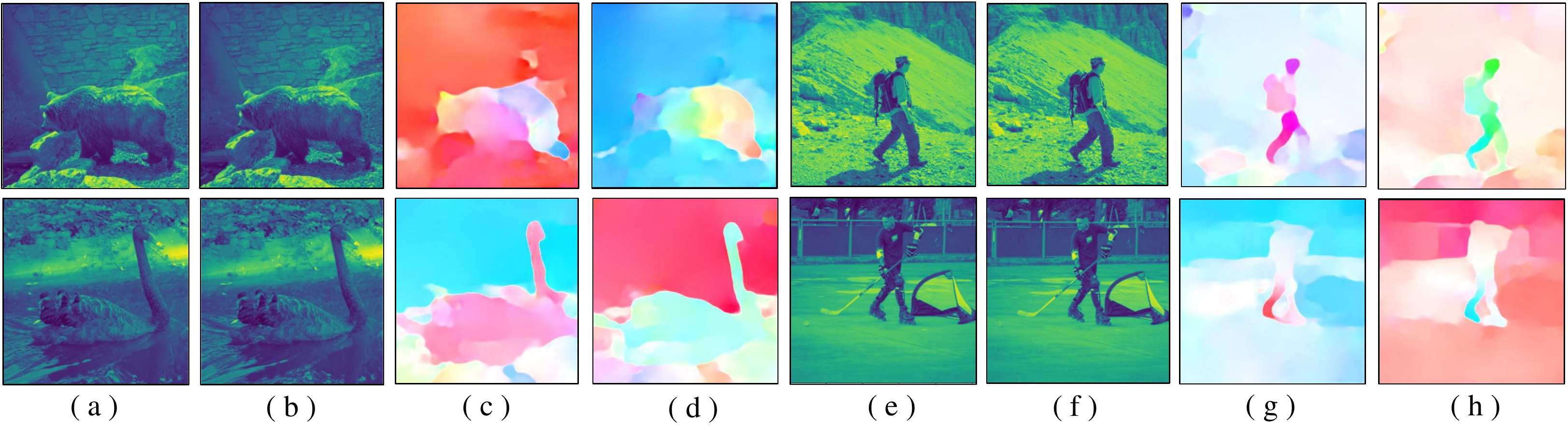}
	\caption{ {Illustration of the fine-tuned optical flow on two simulation datasets.  (a)(b) a pair of consecutive video frames with moving objects of \texttt{Bear\&Blackswan}; (c) the forward optical flow between adjacent frames (a)-(b); (d) the backward optical flow between adjacent frames (b)-(a); (e)(f) a pair of consecutive video frames with moving objects of \texttt{Hike\&Hockey}; (g) the forward optical flow between adjacent frames (e)-(f); (h) the backward optical flow between adjacent frames (f)-(e).}}
	\vspace{-3mm}
	\label{fig:flow}
\end{figure}

\begin{figure*}[th!]
	\centering
	\includegraphics[width=1.95\columnwidth]{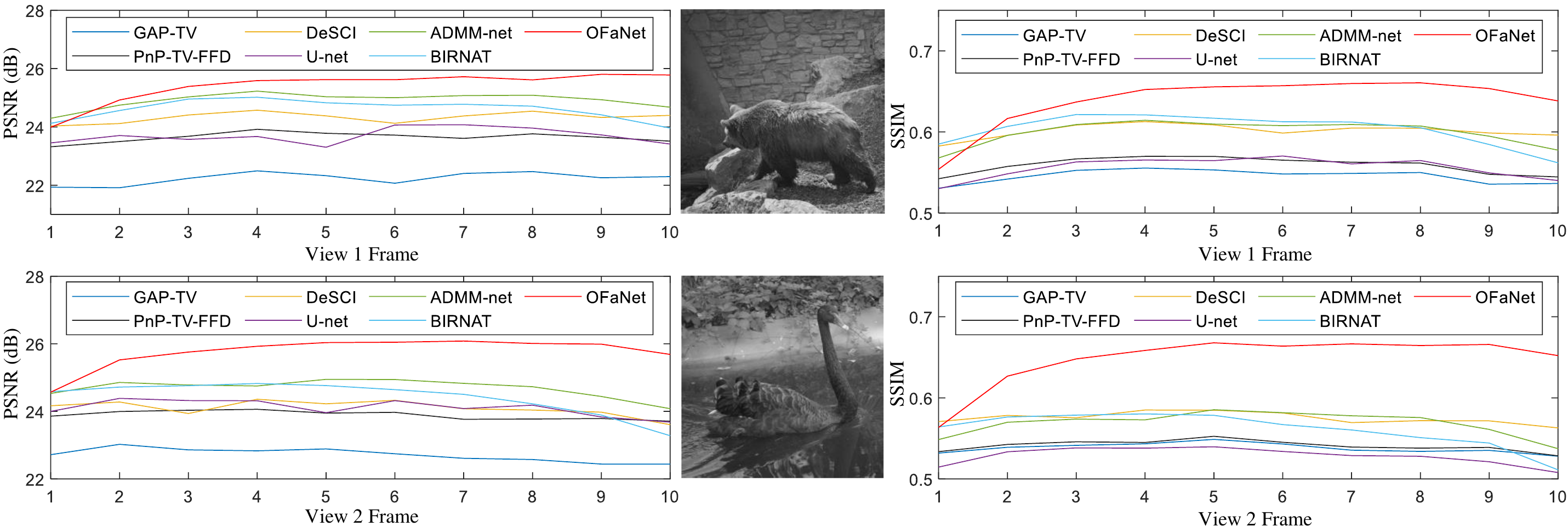}
	\caption{Frame-wise reconstruction quality on PSNR (left) and SSIM (right) by different algorithms on dataset \texttt{Bear\&Blsckswan}. }
	\vspace{-3mm}
	\label{fig:psnr}
\end{figure*}

\paragraph{\textbf{Optical Flow:}} To further investigate the influence of optical flow, we illustrate the flow map on two simulation datasets in Fig.~\ref{fig:flow}.
We can see that a jointly trained optical flow extractor learns informative task-specific features to promote video reconstruction. {It can be observed that the object contours and both the global and local motions are well represented in the optical flow, which gives RNN a helpful guidance for reconstructing the next frame. Both the forward and backward optical flows are very informative and capable to explicitly introduce the dynamic motion into the reconstruction of each adjacent frame, and both the global motions might caused by camera movements and the locally varying motions of various pixel-wise displacement can be detected as shown in the optical flows.
This is beneficial for modeling motions and reducing the artifacts near occlusion boundaries and smoothing the change of moving object across video frames.} Moreover, in the ablation study, we will quantitatively analyze the influence of integrating the information implied in optical flow.

\paragraph{\textbf{Frame-wise Quality:}} {{In order to analyze the reconstruction performance recurrently, we plot the frame-wise PSNR and SSIM curves of different algorithms in Fig.~\ref{fig:psnr}. It can be observed that OFaNet outperforms other counterparts at most of the frames on both PSNR and SSIM. Interestingly, the performance curves of OFaNet seem to follow an up-and-down trend on both PSNR and SSIM, which might due to the recurrent mechanism and the bidirectional optical flow. More specifically, the bidirectional optical flow is degraded into single-direction unavoidably when reconstructing the first and last frames, which results in relatively low quality at these frames and leads to an up-and-down trend. This further demonstrates the effectiveness of our integrated bidirectional optical flow.}}

\begin{table*}[!ht]
\caption{{The average results of PSNR in dB (left entry in each cell) and SSIM (right entry in each cell) and running time per measurement/shot in seconds by \textbf{different variants} of our proposed OFaNet on dual-view simulation datasets.}}
\label{Ablation}
\centering
\resizebox{\textwidth}{!}{
\begin{tabular}{c|c|c|c|c|c|c|c|c}
\hline \hline
Dataset & W/o DA & W/o DNS & W/o RN & W/o OF & W/o Br & W/o JT &DA\&DNS\&BRRN & OFaNet \\ \hline
Bear &24.86 \quad 0.60 & 25.10 \quad 0.61 &24.61 \quad 0.58 &24.70 \quad 0.60 &25.01 \quad 0.61 &25.20 \quad 0.63  &25.27 \quad \textbf{0.64} &\textbf{25.40} \quad \textbf{0.64} \\
Blackswan & 24.97 \quad 0.60 & 25.38 \quad 0.62 &24.57 \quad 0.56 &25.03 \quad 0.60 &25.23 \quad 0.61 &25.48 \quad 0.62  & 25.61 \quad 0.64 &\textbf{25.76} \quad \textbf{0.65} \\ \hline
Boy &25.92 \quad 0.5 5& 25.99 \quad 0.56 &25.59 \quad 0.53 &25.75 \quad 0.55 &25.87 \quad 0.55 &25.98 \quad 0.56  & 26.06 \quad 0.56 &\textbf{26.12} \quad \textbf{0.57} \\
Girl &24.80 \quad 0.64 & 24.96 \quad 0.65 &23.95 \quad 0.60 &25.27 \quad 0.66 &25.36 \quad 0.67 &25.53 \quad 0.67  & 25.61 \quad \textbf{0.68} &\textbf{25.70} \quad \textbf{0.68} \\ \hline
Car-b &22.40 \quad 0.65 & 22.67 \quad 0.66 &21.79 \quad 0.61 &22.18 \quad 0.64 &22.54 \quad 0.65 &22.79 \quad 0.67  & 23.02 \quad 0.68 &\textbf{23.15} \quad \textbf{0.69} \\
Car-w &22.94 \quad 0.69 & 23.25 \quad 0.71 &22.40 \quad 0.67 &23.05 \quad 0.70 &23.39 \quad 0.72 &23.49 \quad 0.72  & 23.63 \quad 0.73 &\textbf{23.73} \quad \textbf{0.74} \\ \hline
Bike &25.60 \quad 0.68 & 26.05 \quad 0.70 &25.03 \quad 0.65 &25.34 \quad 0.67 &25.78 \quad 0.69 &25.85 \quad 0.70  & 26.15 \quad 0.70 &\textbf{26.30} \quad \textbf{0.71}\\
Bus &24.69 \quad 0.77 & 25.28 \quad 0.79 &23.92 \quad 0.73 &24.47 \quad 0.75 &24.93 \quad 0.78 &25.05 \quad 0.78  & 25.17 \quad 0.75 &\textbf{25.43} \quad \textbf{0.79}\\ \hline
Cow &23.12 \quad 0.55 & 23.22 \quad 0.56 &22.67 \quad 0.52 &22.89 \quad 0.54 &23.20 \quad 0.56 &23.08 \quad 0.55  & \textbf{23.18} \quad \textbf{0.55} &{23.15} \quad \textbf{0.55}\\
Dog &25.58 \quad 0.56 & 25.74 \quad 0.57 &25.15 \quad 0.54 &25.65 \quad 0.57 &25.88 \quad 0.58 &25.83 \quad 0.59  & \textbf{25.95} \quad 0.59 &{25.94} \quad \textbf{0.60}\\ \hline
Hike &23.67 \quad 0.56 & 23.81 \quad 0.57 &23.10 \quad 0.51 &23.20 \quad 0.52 &23.46 \quad 0.55 &23.65 \quad 0.55  & 23.80 \quad \textbf{0.57} &\textbf{23.85} \quad \textbf{0.57}\\
Hockey &26.01 \quad 0.76 & 26.60 \quad 0.78 &25.29 \quad 0.73 &25.52 \quad 0.74 &26.31 \quad 0.77 &26.48 \quad 0.77  & 26.63 \quad 0.78 &\textbf{26.85} \quad \textbf{0.79}\\ \hline\hline
Average &24.55 \quad 0.63 & 24.83 \quad 0.65 &24.01 \quad 0.60 &24.42 \quad 0.63 &24.75 \quad 0.65 &24.87 \quad 0.65  &25.01 \quad 0.66 &\textbf{25.12} \quad \textbf{0.67} \\
Time(s) &0.4829 &0.4862 &\textbf{0.0057} &0.0861 &0.3013 &0.4898 &0.9067 &0.4893\\ \hline \hline
\end{tabular}}
\end{table*}

\paragraph{\textbf{Ablation Study:}}
To demonstrate the effectiveness of each module in OFaNet, we conduct experiments on the six dual-view simulation datasets using various versions of OFaNet, as listed in Table~\ref{Ablation}, where the abbreviations of different variants of OFanet are shown in the first row, specifically corresponding to:
\begin{itemize}
\item {\textbf{OFaNet without diversity amplifier (W/o DA)} excludes the diversity amplified images $\Dmat_1,\Dmat_2,\Dmat_3,\Dmat_4$ from the entire reconstruction framework. Thus, the input of `Dual-net Separator' is changed to $\small \mathbf{DNS}_1^0=[\overline{\Ymat}, \overline{\Ymat} \odot \Cmat_{1}^1, \overline{\mathbf{Y}} \odot \Cmat_{1}^2, \ldots,$ $\overline{\mathbf{Y}} \odot \Cmat_{1}^B]_{c3}$ and $\small \mathbf{DNS}_2^0=\left[\overline{\Ymat}, \overline{\mathbf{Y}} \odot \Cmat_{2}^1, \overline{\mathbf{Y}} \odot \Cmat_{2}^2, \ldots, \overline{\mathbf{Y}} \odot \Cmat_{2}^B\right]_{c3}$, and the input $\widehat{\mathcal{F}_D}(\Dmat_{1:4})$ in \eqref{bidirection_rnn} is removed from the bidirectional refining network. As shown in Table \ref{Ablation}, the diversity amplifier leads to average improvements both on PSNR (0.57) and SSIM (0.04) of six datasets with neglectable computational cost. This demonstrates the effectiveness and efficiency of diversity amplifier for separating and reconstructing dual videos.}

\item {\textbf{OFaNet without dual-net separator (W/o DNS)} is utilized to evaluate the effectiveness of the dual-branches framework. Considering the optical flow can only be well estimated based on the coarse reconstruction results, it is not appropriate to directly remove the dual-net separator. Thus, we replace the dual-branches network with the single-branch network with the same architecture but the different output channel at the last layer. To be more specific, for the original dual-net separator, we employ two network without sharing parameters to output each video respectively, while the variety `OFaNet W/o Dual-net Separator' utilizes a single network to output two videos as a single sequential data cube. Mathematically, the single-branch network can be expressed as:
    {
    \begin{equation}
    \label{eq_single}
    \small
    \begin{array}{l}
    \widetilde{\Xmat}=\mathcal{F}^{cnn3}([\mathbf{DNS}^2,\mathbf{DNS}^1]_{c3}),\\
    \mathbf{DNS}^2=\mathcal{F}^{cnn2}(\mathbf{DNS}^1),\\
    \mathbf{DNS}^1=\mathcal{F}^{cnn1}(\mathbf{DNS}^0),\\
    \mathbf{DNS}^0=[\overline{\Ymat}, \Dmat_1, \Dmat_2, \Dmat_3, \Dmat_4,
    \overline{\Ymat} \odot \Cmat_{1}^1, \overline{\mathbf{Y}} \odot \Cmat_{1}^2, \\
    \ldots, \overline{\mathbf{Y}} \odot \Cmat_{1}^B, \overline{\mathbf{Y}} \odot \Cmat_{2}^1, \overline{\mathbf{Y}} \odot \Cmat_{2}^2, \ldots, \overline{\mathbf{Y}} \odot \Cmat_{2}^B ]_{c3}
    ,\\
    \end{array}
    \end{equation}}
where {the output} $\widetilde{\Xmat} = [\widetilde{\Xmat}_1, \widetilde{\Xmat}_2]_{c3}$ is the concatenation of two reconstructed videos. As we can see from Table \ref{Ablation}, the single-branch network results in average descent on PSNR (0.29) and SSIM (0.02) compared to the dual-net separator. Intuitively, the dual-branches network is more appropriate to reconstruct individual video by respectively taking its corresponding coding patterns into consideration, while the computational cost is marginal and acceptable (0.0031s).}

\item \textbf{OFaNet without refine net (W/o RN)} only contains the diversity amplifier and the dual-net separator. {On the one hand}, it can be seen that this {variety} provides very limited results {compared with OFaNet}, which implies the refining network is beneficial to the reconstruction performance, {leading} to an average improvement on PSNR (1.11) and SSIM (0.07) of six simulated datasets.
{On the other hand}, the reconstruction results of this variety show superiority compared with the basic model U-net, which could verify the effectiveness of our proposed diversity amplifier and dual-net separator. {It is worth noticing that this architecture {without refine net} only cost less than 0.01s for testing, which is applicable with the fast inference speed and a comparable reconstruction performance for the applications with strict time demand such as self-driving systems and motion control.}

\item \textbf{OFaNet without optical flow (W/o OF)} excludes the optical flow from our proposed OFaNet, which can be obversed from Table~\ref{Ablation} that the performance degrades {severely (0.7 on PSNR and 0.04 on SSIM)} compared with OFaNet. The significant contribution of optical flow is verified, where the optical flow helps the reconstruction of each video by feeding the motion information as a guidance.
{Intuitively, the dynamic motion provides the information to guide the refining network to estimate the output according to the predicted motion, leading to smooth results across time by focusing on the locally varying dynamic object and the global movement in the background.}

\item \textbf{OFaNet without bidirection (W/o Br)} only includes the forward optical flow $F^{t \rightarrow t+1}$ without the backward optical flow $F^{t+1 \rightarrow t}$, and we can see that adding the backward optical flow is beneficial to video reconstruction. The bidirectional dynamic information is explored among adjacent frames both forwardly and backwardly, which can further improve the quality of reconstruction by integrating into RNN.

\item {\textbf{OFaNet without joint training (W/o JT)} utilizes the pre-trained model of ~\cite{IlgMSKDB17} and initializes the weights of Flownet from the released pre-trained model \footnote{https://github.com/NVIDIA/flownet2-pytorch/} without fine-tuning. We can see from Table~\ref{Ablation} that joint training the parameters of the whole framework with mean square error (MSE) loss is beneficial to video reconstruction, which helps the extracted optical flow to better match the video SCI problem.}

\begin{table*}[!ht]
\caption{{The average results of PSNR in dB (left entry in each item), SSIM (right entry in each item) and running time per measurement/shot in seconds by different algorithms on dual-view simulation datasets with {\bf different compression rates}.}}
\label{Frames}
\centering
\resizebox{0.8\textwidth}{!}{
\begin{tabular}{c|c|c|c|c||c|c|c|c}
\hline
\multicolumn{2}{c|}{\multirow{2}{*}{Algorithm}} &\multicolumn{3}{c|}{Optimization based methods}  &\multicolumn{4}{c}{Deep learning based methods} \\ \cline{3-9}
\multicolumn{2}{c|}{} & GAP-TV  & PnP-TV-FFD  & DeSCI & U-net & ADMMnet & DA\&DNS & OFaNet   \\ \hline
\multirow{3}{*}{\bf B=6}
& PSNR &23.93 &21.73 &26.17 &23.39 &26.05&25.55 &\textbf{26.50}\\
& SSIM &0.65 &0.60 &0.75 &0.62 &0.71&0.68 &\textbf{0.73}\\
& Time &11.8 &49.5 &17993.5 &0.0214 & 0.0667 &\textbf{0.0052} &0.3766\\ \hline
\multirow{3}{*}{\bf B=10}
& PSNR &22.49 &22.54 &24.51 &22.97 &24.14 &24.01 &\textbf{25.12}\\
& SSIM &0.60 &0.57 &0.64 &0.55 &0.60 &0.60 &\textbf{0.67}\\
& Time &16.9 &52.3 &21600.4 &0.0216 &0.0680 &\textbf{0.0057} &0.4893\\ \hline
\multirow{3}{*}{\bf B=14}
& PSNR &21.39 &20.69 &21.73 &22.48 &23.18 &23.16 &\textbf{24.25} \\
& SSIM &0.54 &0.52 &0.50 &0.55 &0.55 &0.56 &\textbf{0.63}\\
& Time &28.7 &90.0 &60636.6 &0.0215 &0.0689 &\textbf{0.0058} &0.9481\\ \hline
\end{tabular}}
\end{table*}

\begin{table*}[!ht]
\caption{{The average results of PSNR in dB (left entry in each item) and SSIM (right entry in each item) and running time per measurement/shot in seconds by different algorithms on dual-view simulation datasets with {\bf different level Gaussian noise}.}}
\label{Noise}
\centering
\resizebox{\textwidth}{!}{
\begin{tabular}{c|c|c|c||c|c|c|c}
\hline \hline
\multirow{2}{*}{Noise Level $\sigmav$} &\multicolumn{3}{c|}{Optimization based methods}  &\multicolumn{4}{c}{Deep learning based methods} \\ \cline{2-8}
& GAP-TV  & PnP-TV-FFD  &  DeSCI  & U-net  & ADMMnet  & BIRNAT & OFaNet   \\ \hline
0 &22.49 \quad 0.60 &22.54 \quad 0.57 &24.51 \quad 0.64 &22.97 \quad 0.55 &24.14 \quad 0.60 &23.93 \quad 0.61 &25.12 \quad 0.67\\
0.01 &21.69 \quad 0.53 &20.54 \quad 0.51 & 22.79 \quad 0.56 &22.88 \quad 0.54 &23.76 \quad 0.58&23.45 \quad 0.59 &24.83 \quad 0.65\\
0.05 &16.73 \quad 0.26 &13.92 \quad 0.29 &15.54 \quad 0.35 &19.32 \quad 0.41& 20.08 \quad 0.35 &19.71 \quad 0.44 &20.69 \quad 0.47 \\
0.1 &13.12 \quad 0.12 &9.94 \quad 0.16 &8.65 \quad 0.14 &16.12 \quad 0.26 &16.36 \quad 0.18 &16.07 \quad 0.33 &17.09 \quad 0.34 \\
0.2 &9.16 \quad 0.05 &8.01 \quad 0.08 &5.35 \quad 0.23&12.40 \quad 0.16 & 12.22 \quad 0.07 &13.07 \quad 0.24 &14.83 \quad 0.26\\  \hline \hline
\end{tabular}}
\end{table*}

\item {\textbf{Diversity Amplifier \texttt{+} Dual-net Separator \texttt{+} Bidirectional Recurrent Refining Network \\ (DA\&DNS\&BRRN)} combines the "OFaNet W/o refine net" and a bidirectional recurrent network (a forward and a backward rnn) together for video reconstruction.
As shown in Table~\ref{Ablation}, the reconstruction performance is comparable or even better than OFaNet, but with significantly more computational cost. In specific, both the memory and testing time in each epoch of "Bidirectional Recurrent Refining Network" is twice larger than OFaNet, and its training time is much longer than OFaNet. Bigger network parameters also make the model difficult to train and converge. The proposed OFaNet condenses the bidirectional network to save half of the required memory and meanwhile makes full use of the bidirectional motion information.}

\end{itemize}
{In terms of the above oblation study, we can summarize that the final version of OFaNet results in the best performance according to all these above varieties.}

\paragraph{{\textbf{Results of different compression rates:}}}
{In order to evaluate the performance of the proposed algorithm dealing with different compression rates, we show the results on six simulation data with compression rates $B=\{6, 10, 14\}$ for each view. As shown in Table~\ref{Frames}, the proposed OFaNet outperforms its counterparts across different compression rates. {DeSCI performs well across small compression rates ($B=\{6, 10\}$) but results in low quality with the big compression rate ($B=14$), which may due to the parameter setting.} Moreover, DeSCI takes about 16 hours to reconstruct dual videos with the size of $256 \times 256 \times 14$ from a single measurement, which is difficult to support practical applications. In contrast, the proposed OFaNet achieves state-of-the-art performance within 1 second for fast inference. It is worth noting that the variety {DA\&DNS} (i.e. OFaNet without refine net) of the proposed model achieves descent performance in the shortest time ($\le$ 0.006s), which is a computationally efficient method. For applications with strict time demand such as self-driving systems~\cite{Lu20SEC} and motion control, the variety network {DA\&DNS} is a better choice to reconstruct videos with reasonably good quality within very fast time.}

\begin{table*}[htbp!]
  \caption{{The average results of PSNR in dB (left entry in each item), SSIM (right entry in each item) and running time per measurement/shot in seconds by different algorithms on six {\bf single-view} grayscale benchmark simulation datasets. The best results are \textbf{bold}, and the second best results are \underline{underlined}.}}
  \label{Table:Sim-single}
  \centering
  \resizebox{\textwidth}{!}{
  \begin{tabular}{c|c|c|c|c|c|c|c|c}
    \hline \hline
    Algorithm & \texttt{Kobe} & \texttt{Traffic} &\texttt{Runner} &\texttt{Drop} &\texttt{Aerial} &\texttt{Vehicle} &\texttt{Average} & Time\\
    \hline
    GAP-TV &26.45 \quad 0.845 &20.89 \quad 0.715  &28.81 \quad 0.909 &34.74 \quad 0.970 &25.05 \quad 0.828  &24.82 \quad 0.838 &26.79 \quad 0.858 &4.2\\
    DeSCI  &\textbf{33.25} \quad \textbf{0.952}  &{28.72} \quad 0.925  &\textbf{38.76} \quad {0.969}  &\textbf{43.22} \quad \textbf{0.993}  &25.33 \quad 0.860  &27.04 \quad 0.909  &32.72 \quad 0.935 & 6180\\
    U-net & 29.02 \quad 0.861 & 23.45 \quad 0.838 & 34.43 \quad 0.958 & 36.77 \quad 0.974 & 27.52 \quad 0.882 & 26.40 \quad 0.886 & 29.26 \quad 0.900 & \textbf{0.023}\\
    ADMM-net &31.41 \quad 0.936 &27.40 \quad 0.918 &37.18 \quad 0.971 &40.71 \quad 0.989 &28.43 \quad 0.908 &27.43 \quad 0.922 &32.09 \quad 0.941 &\underline{0.067} \\
    BIRNAT &\underline{32.71} \quad \underline{0.950} &\textbf{29.33} \quad \textbf{0.942} &\underline{38.70} \quad \textbf{0.976} &\underline{42.28} \quad \underline{0.992} &\textbf{28.99} \quad \textbf{0.927} &\textbf{27.84} \quad \textbf{0.927} &\textbf{33.31} \quad \textbf{0.951} &0.16\\ \hline
    OFaNet &32.29 \quad 0.942 &\underline{28.76} \quad \underline{0.926} &38.37 \quad \underline{0.970} &41.84 \quad 0.990 &\underline{28.46} \quad \underline{0.905} &\underline{27.51} \quad \underline{0.919} &\underline{32.88} \quad \underline{0.942} &0.14\\
    \hline \hline
  \end{tabular}}
\end{table*}

\begin{figure}[th!]
	\centering
	\includegraphics[width=1.0\columnwidth]{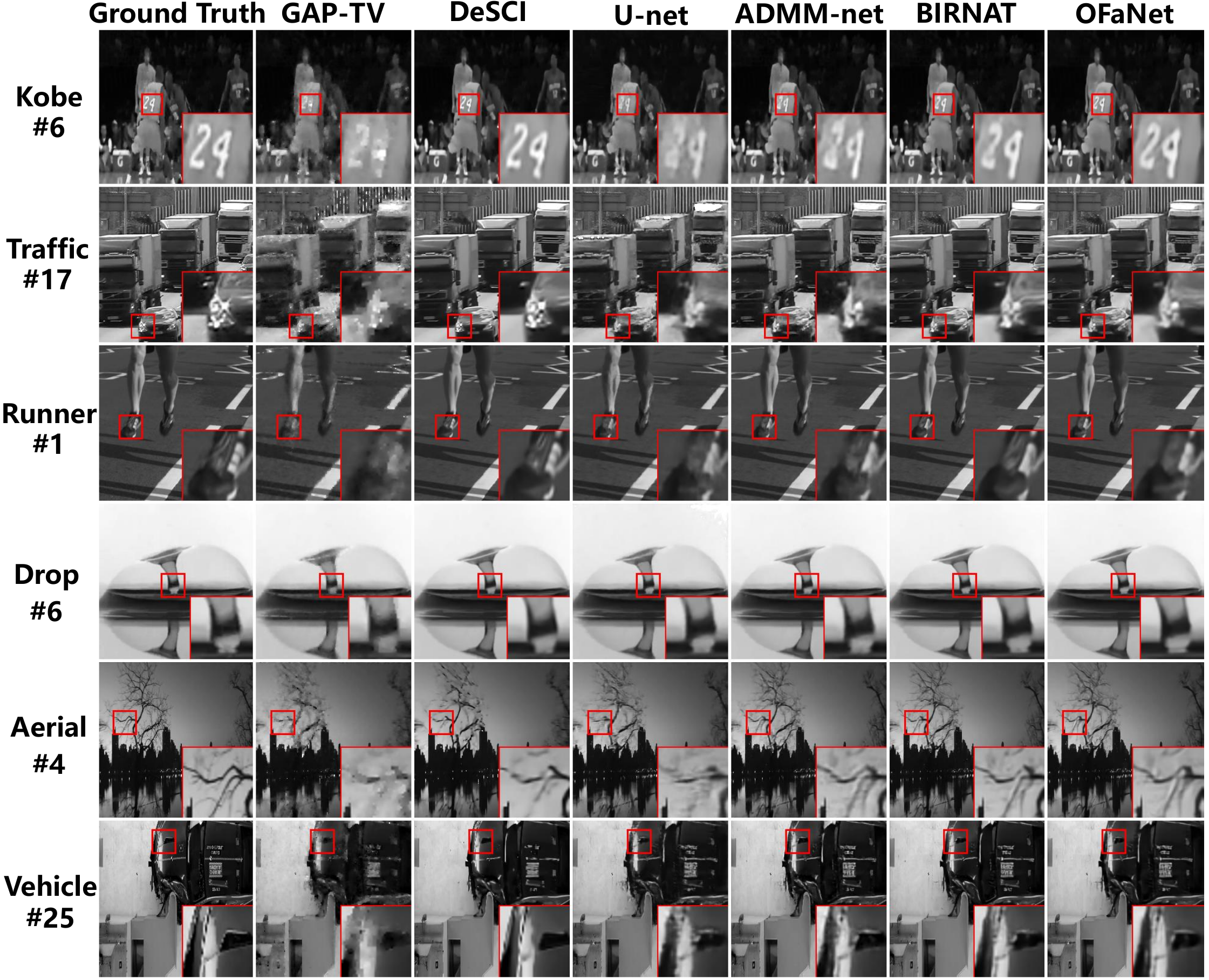}
	\caption{Selected reconstruction frames of six single-view grayscale benchmark simulation datasets.}
	\vspace{-3mm}
	\label{fig:Sim-single}
\end{figure}

\paragraph{\textbf{Robustness to Gaussian Noise:}}
{In order to evaluate the robustness of the proposed OFaNet, we investigate the effect of Gaussian noise corresponding to different algorithms. Before adding noise to simulation data, all the measurements are normalized to [0, 1]. Then the zero-mean Gaussian noise is added to the measurements with standard deviation $\sigma$ of \{0, 0.01, 0.05, 0.1, 0.2\}, respectively. For the optimization based algorithms, we retrain the entire framework from scratch. For the deep learning based methods, we directly test on the noisy test data using the previously well-trained parameters. As shown in Table~\ref{Noise}, four deep learning based methods are more robust and stable on the noisy data compared with the optimization based methods. In specific, the performance of the runner-up DeSCI drops very fast when the measurement is corrupted (fine tuning the parameters might be able to provide better results but time consuming). Surprisingly, the competitive deep learning method ADMM-net performs even worse than U-net, while the BIRNAT shows robustness on the corrupted data, which might due to its powerful representative capability implied in the huge network and sequential modeling. The performance of the OFaNet degrades slower than others, which indicates the proposed model is more robust to noise and can relieve the effect of Gaussian noise.}

\paragraph{{\textbf{Results on Single-view Simulated SCI Data:}}}

{In order to evaluate the proposed method in a broader scope, we further demonstrate the experiments on the single-view system using six widely used benchmark simulated data, i.e. \texttt{Kobe, Runner, Drop, Traffic, Aerial} and \texttt{Vehicle}. The modification of OFaNet is small to extend to single-view SCI, where we remove the diversity amplifier and only keep one branch of the dual-net separator, and utilize the same training data and experiments setting as used in BIRNAT~\cite{Cheng20ECCV_BIRNAT}. The results of different comparison methods are given in Table~\ref{Table:Sim-single}. It can be seen that the proposed OFaNet achieves competitive results, 0.16dB in PSNR higher than the strong baseline DeSCI with about 44,000$\times$ shorter testing time. Compared to state-of-the-art method BIRNAT, the proposed OFaNet occupies 2 times lower memory (8934MB) during training than BIRNAT (17748MB), which is important because the big GPU memory consumption will preclude the practical large-scale SCI applications. Fig.~\ref{fig:Sim-single} plots selected reconstruction frames of different algorithms on six datasets. It can be observed that OFaNet achieves comparable visual results compared with the state-of-the-art methods; clean and sharp contours and fine details can be provided by the proposed OFaNet.}

\begin{figure}[ht!]
	\centering
	\includegraphics[width=1.0\columnwidth]{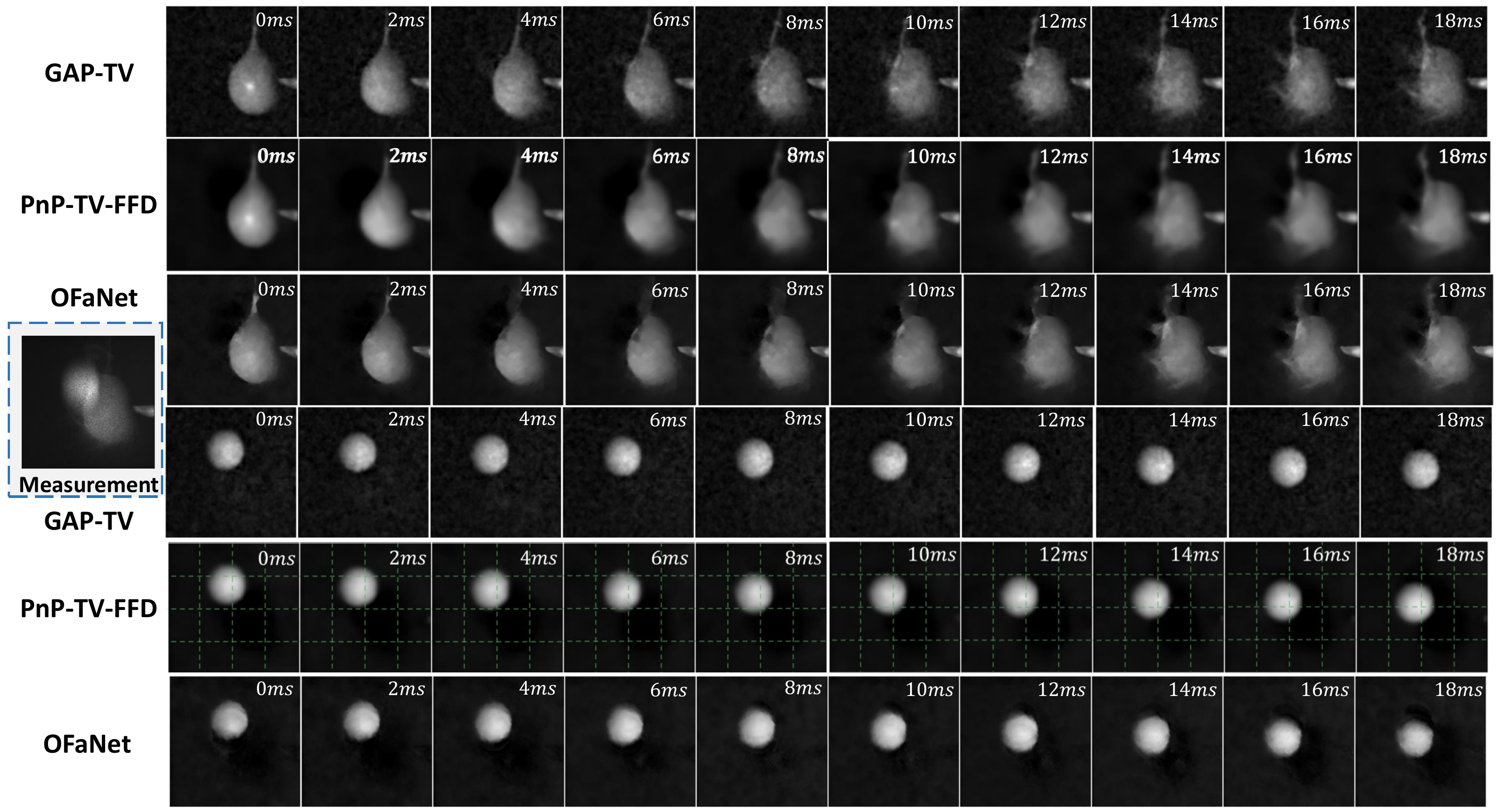}
	\vspace{-3mm}
	\caption{Reconstruction results by different algorithms for real SSTCI data  \texttt{popping water balloon punctured by a knife} (top three rows) and \texttt{a falling Ping-Pong ball} (bottom three rows).}
	\label{fig:real1_full}
\end{figure}

\begin{figure}[ht!]
	\centering
	\includegraphics[width=1.0\columnwidth]{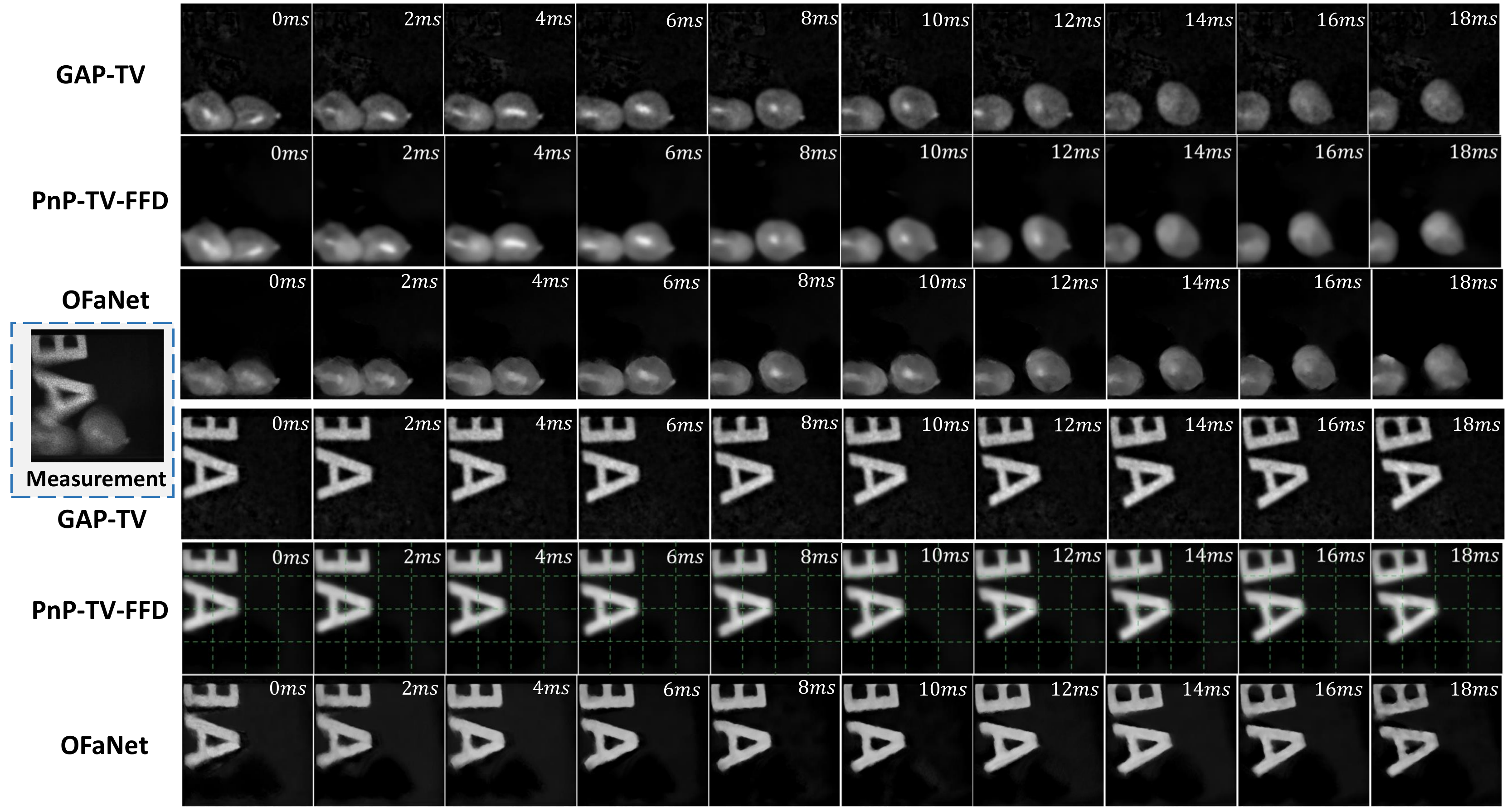}
	\vspace{-3mm}
	\caption{Reconstruction results by different algorithms for real SSTCI data  \texttt{colliding water balloons} (top three rows) and \texttt{flying letters} (bottom three rows).}
	\label{fig:real2_full}
\end{figure}

\subsection{Results on Real SCI Data}
Lastly, we apply the proposed OFaNet to real data captured by the SSTCI camera.
Following the same setting in~\cite{qiao2020snapshot}, two FoV videos are encoded into one snapshot measurement.
Three datasets are utilized here. The first one is \texttt{a popping water balloon punctured by a knife} and \texttt{a falling Ping-Pong ball}, {with the size of $650 \times 650 \times 10$ corresponding to each view.} The second dataset is composed of \texttt{two colliding water balloons and two flying letters} ($650 \times 650 \times 10$ for each view). The third video {contains} \texttt{5 pendulum balls and 4 falling dominoes}, where each FoV video has the size of $650 \times 650 \times 20$. The real captured datasets capture high-speed videos in our daily life {with} unavoidable noise inside and thus are more challenging to reconstruct.
The camera was working at 50 frames per second (fps) during capturing these measurements and thus each measurement corresponds to 20 ms in real life. Therefore, when 10 frames of each view are reconstructed from the single measurement, each frame lasts 2 ms in real life; this is 500fps high speed video. Similarly, when 20 frames of each view are reconstructed from the single measurement, each frame lasts 1 ms in real life and the reconstructed video is of 1000fps.

The measurements and the corresponding reconstructed frames of two real dual-view SCI system with $B=10$ are demonstrated in Fig.~\ref{fig:real1_full} and Fig.~\ref{fig:real2_full}, respectively, with full videos shown in the SM. It can be observed that in both cases, our OFaNet can provide better or at least competitive results with a significant saving on the computational time compared to existing algorithms. The reconstructed videos by GAP-TV have unpleasant artifacts, and PnP-TV-FFD shows blurry boundaries and {over-smoothness}. PnP-TV-FFD is capable of providing better results on real data compared to the performances on simulation datsets, which might due to the FFD denoiser is more appropriate for real datasets with unavoidable noise. It can also be seen that OFaNet provides sharp boundaries and fine details with less artifacts and fuzziness from the single compressed measurement.

\begin{figure}[ht!]
	\centering
	\includegraphics[width=1.0\columnwidth]{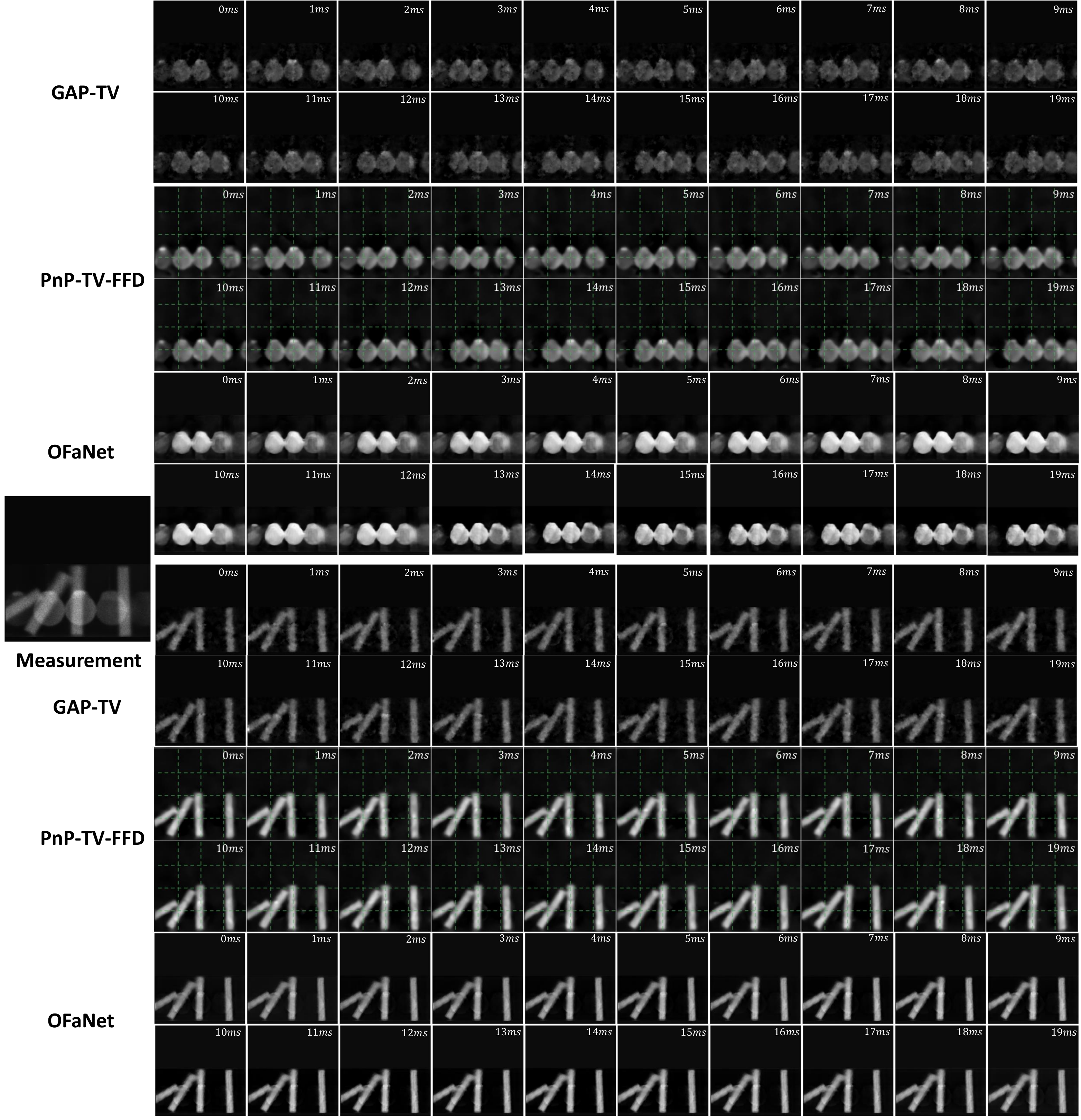}
	\caption{ Reconstruction results using different algorithms for the real SSTCI data with $B=20$:  \texttt{pendulum balls} (top six rows) and \texttt{falling Ping-Pong balls} (bottom six rows).}
	\vspace{-3mm}
	\label{fig:real3_full}
\end{figure}

{Another snapshot measurement of \texttt{pendulum balls and falling dominoes} with $B=20$ and the corresponding reconstructed results are shown in Fig.~\ref{fig:real3_full}, with full videos shown in the SM. {It can be observed that GAP-TV produces significant noise and unclear boundaries, while PnP-TV-FFD tends to over smooth the moving object. OFaNet is capable of providing relatively clear and distinct pendulum balls, as well as the sharp and straight contours of falling dominoes.}}

In summary, the reconstruction results of our OFaNet are of higher quality compared with other algorithms, and the inference of our algorithm is significantly faster. This indicates the applicability and efficiency of our proposed OFaNet in the real SSTCI systems.

\begin{figure}[ht!]
	\centering
	\includegraphics[width=1.0\columnwidth]{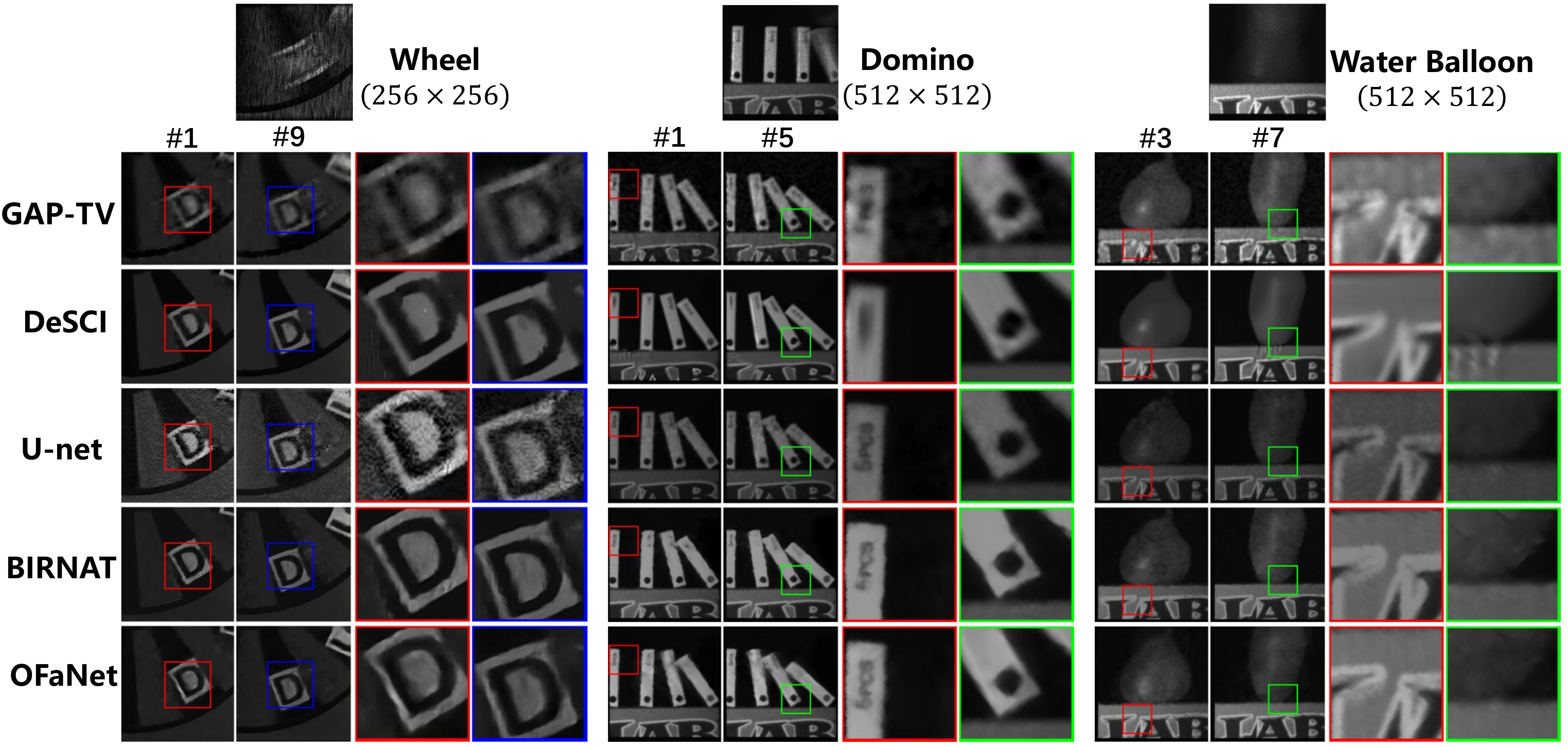}
	\caption{ The reconstructed frames of three single-view real data \texttt{Wheel}, \texttt{Domino}, and \texttt{Water Balloon}.}
	\vspace{-3mm}
	\label{fig:single_real}
\end{figure}

\paragraph{\textbf{Results on Single-view Real SCI Data:}}
{To evaluate the effectiveness of OFaNet on real applications of the single-view SCI system, we conduct experiments on three real data captured by the SCI cameras~\cite{Patrick13OE,Qiao2020_APLP}. For the snapshot measurement \texttt{Wheel}, we recover a $256\times256\times14$ high-speed video. In addition, the larger scale snapshot measurements \texttt{Domino} and \texttt{Water Balloon} are recovered as two videos of size $512\times12\times10$. As shown in Fig.~\ref{fig:single_real}, it can be seen that GAP-TV introduces unpleasant noise; DeSCI over smooth the details such as the letters in \texttt{domino}; U-net results in blurry edges and more noise such as in the letter `D' of \texttt{Wheel}; BIRNAT provides finer details and sharper edges compared with other methods; OFaNet provides relatively clear contours with fewer artifacts and less noise, even though it results in some blurry parts. However, the proposed OFaNet utilizes much less testing time than DeSCI and twice lower GPU memory compared with BIRNAT. These experiments indicate both the applicability and effectiveness of the proposed algorithm in real applications.}

\section{Conclusions}
We have proposed an efficient deep learning network for the reconstruction of dual-view video snapshot compressive imaging, which implements joint field-of-view and temporal compressive sensing, shedding light to high throughput machine vision systems. Inspired by the hardware encoding principle, we develop a diversity amplifier to enhance the differences of the scenes from two FoVs, and design a dual-net separator to reconstruct two views from the single measurement. Following this, we integrate recurrent mechanism and optical flow into our reconstruction network to achieve competitive results in a short time. Though the network is proposed for dual-view systems, it can be extended to single-view systems and we believe it can work for multi-view systems with moderate modifications.
This will pave the way of real applications of SCI systems on robots, self-driving vehicles, \etc


\bibliographystyle{plain}      
\bibliography{dual-view_arxiv}

\end{document}